\documentclass[12pt,preprint]{aastex}

\slugcomment{}

\shorttitle{New Fe~{\sc ii} Abundance Scale}
\shortauthors{Kraft \& Ivans}

\begin{document}
\topmargin -0.3in
\textheight 10in
\textwidth 7in


\title{A Globular Cluster Metallicity Scale Based on the Abundance of Fe~{\sc ii}}


\author{Robert P.\ Kraft}
\affil{UCO/Lick Observatory, University of California, Santa Cruz, CA 95064}
\email{kraft@ucolick.org}

\and

\author{Inese I.\ Ivans\altaffilmark{1,2}}
\affil{Dept.\ of Astronomy and McDonald Observatory, The University of Texas at Austin,  Austin, TX 78712}
\email{}


\altaffiltext{1}{present address: Astronomy Department, California Institute of Technology, Mail Stop 105-24, Pasadena, CA 91125; iii@astro.caltech.edu}
\altaffiltext{2}{Hubble Fellow}


\begin{abstract}
      Assuming that in the atmospheres of low-mass, metal-poor 
red giant stars, one-dimensional models based on local 
thermodynamic equilibrium accurately predict the abundance of 
iron from Fe~{\sc ii}, we derive a globular cluster metallicity 
scale based on the equivalent widths of Fe~{\sc ii} lines 
measured from high resolution spectra of giants in 16 key 
clusters lying in the abundance range --2.4 $<$ 
[Fe/H]$_{\rm II}$ $<$ --0.7.  We base the scale largely on the 
analysis of spectra of 149 giant stars in 11 clusters by the 
Lick-Texas group supplemented by high-resolution studies of 
giants in five other clusters.   We also derive {\it ab initio} 
the true distance moduli for certain key clusters (M5, M3, M13, 
M92, and M15) as a means of setting stellar surface gravities.
Allowances are made for changes in the abundance scale if one
employs: (1) Kurucz models with and without convective 
overshooting to represent giant star atmospheres in place of 
MARCS models and (2) the Houdashelt {\it et al.}\ 
color-temperature scale in place of the Alonso {\it et al.}\ 
scale. 

      We find that [Fe/H]$_{\rm II}$ is correlated linearly 
with $<$$W'$$>$, the reduced strength of the near infra-red 
Ca~{\sc ii} triplet defined by Rutledge {\it et al.}, 
although the actual correlation coefficients depend on the 
atmospheric model employed.  The correlations, limited to the
range --2.4 $<$ [Fe/H]$_{\rm II}$ $<$ --0.7 are as follows:
\begin{enumerate}
\item 
\indent
[Fe/H]$_{\rm II}$ = 0.531 $\times$ $<$$W'$$>$ -- 3.279 (MARCS)
\item 
\indent
[Fe/H]$_{\rm II}$ = 0.537 $\times$ $<$$W'$$>$ -- 3.225 (Kurucz with conv.\ ove
rshooting)
\item 
\indent
[Fe/H]$_{\rm II}$ = 0.562 $\times$ $<$$W'$$>$ -- 3.329 (Kurucz without conv.\ 
overshooting)
\end{enumerate}

      We also discuss how to estimate [X/Fe]-ratios. We suggest 
that C, N, and O, as well as elements appearing in the spectrum 
in the singly ionized state, {\it e.g.}, Ti, Sc, Ba, La and Eu, 
should be normalized to the abundance of Fe~{\sc ii}. Other 
elements, which appear mostly in the neutral state, but for 
which the dominant species is nevertheless the ionized state, 
are probably best normalized to Fe~{\sc i}, but uncertainties 
remain.

\end{abstract}





\section{Introduction\label{1intro}}

Attempts to establish an accurate metallicity scale for globular clusters 
have peppered the astronomical literature ever since it was first 
conclusively established that M92 had an [Fe/H]-ratio\footnote{
We adopt the usual spectroscopic notations that
[A/B] $\equiv$ {\rm log}$_{\mathrm 10}$(N$_{\mathrm A}$/N$_{\mathrm B}$)$_{\star}$ -- {\rm log}$_{\mathrm 10}$(N$_{\mathrm A}$/N$_{\mathrm B}$)$_{\odot}$, and that {\rm log} $\epsilon$(A) $\equiv$ {\rm log}$_{\mathrm 10}$(N$_{\mathrm A}$/N$_{\mathrm H}$) + 12.0, for elements A and B.}
two orders of magnitude below that of the Sun \citep{hwg59}.  
Even the most luminous giants in globular clusters appear as 
relatively faint objects, and thus are difficult to study at 
high spectral resolution.  Thus astronomers have generally 
employed metallicity-sensitive photometric or low-to-medium 
resolution spectroscopic indices so that observations could 
be made of a large sample of clusters distributed widely in 
the Galaxy. 

The first truly extensive investigation of this kind (Zinn 
\& West 1984\nocite{zw84}; hereafter ``ZW84'') was based on 
metallicity-sensitive photometric indices of integrated 
cluster light. The most recent version (Rutledge 
{\it et al.}\ 
1997a\nocite{rutledge+97a},b\nocite{rutledge+97b}) employed 
equivalent width measurements (``EWs'') of the strong near 
infra-red (``IR'') Ca~{\sc ii} triplet lines in low 
resolution spectra of individual cluster giants. The method 
has the advantage that it is relatively insensitive to 
errors in cluster reddening and distance modulus. However, 
regardless of the technique employed, these investigations 
share a common calibration problem: giants in a small group 
of relatively nearby clusters must be subjected to high 
resolution abundance analysis so that the photometric or 
spectroscopic indices can be calibrated in terms of [Fe/H]. 
Metallicity estimates for the vast majority of clusters are 
then obtained by interpolating the correlation of the 
observed values of the indices and thus calculating the 
value of [Fe/H]. The metallicity scale is therefore 
dependent on the reliability of [Fe/H]-determinations for 
certain relatively nearby key clusters.

At the present time, two competing calibrations exist, 
based on high resolution spectroscopic analysis. The older 
version, developed in basic papers by \citet{sz78}, 
\citet{zinn80}, and ZW84\nocite{zw84}, employed the 
pioneering high resolution studies of \citet{cohen78, 
cohen79, cohen80, cohen81}, who observed and 
analyzed small samples of bright giants in a few key 
clusters.  These results were supplemented by Butler's 
(1975)\nocite{butler75} measurements of the Preston 
(1959\nocite{preston59}, 1961\nocite{preston61}) $\Delta$$S$ 
determinations for RR Lyraes in several clusters, and his 
[Fe/H]-calibration of $\Delta$$S$ based on high resolution 
spectroscopic observations of nearby RR Lyraes in the halo 
field.\footnote{
The most recent update of the $\Delta$$S$ calibration is that 
of \citet{clementini+95}.}

\begin{figure}
\plotone{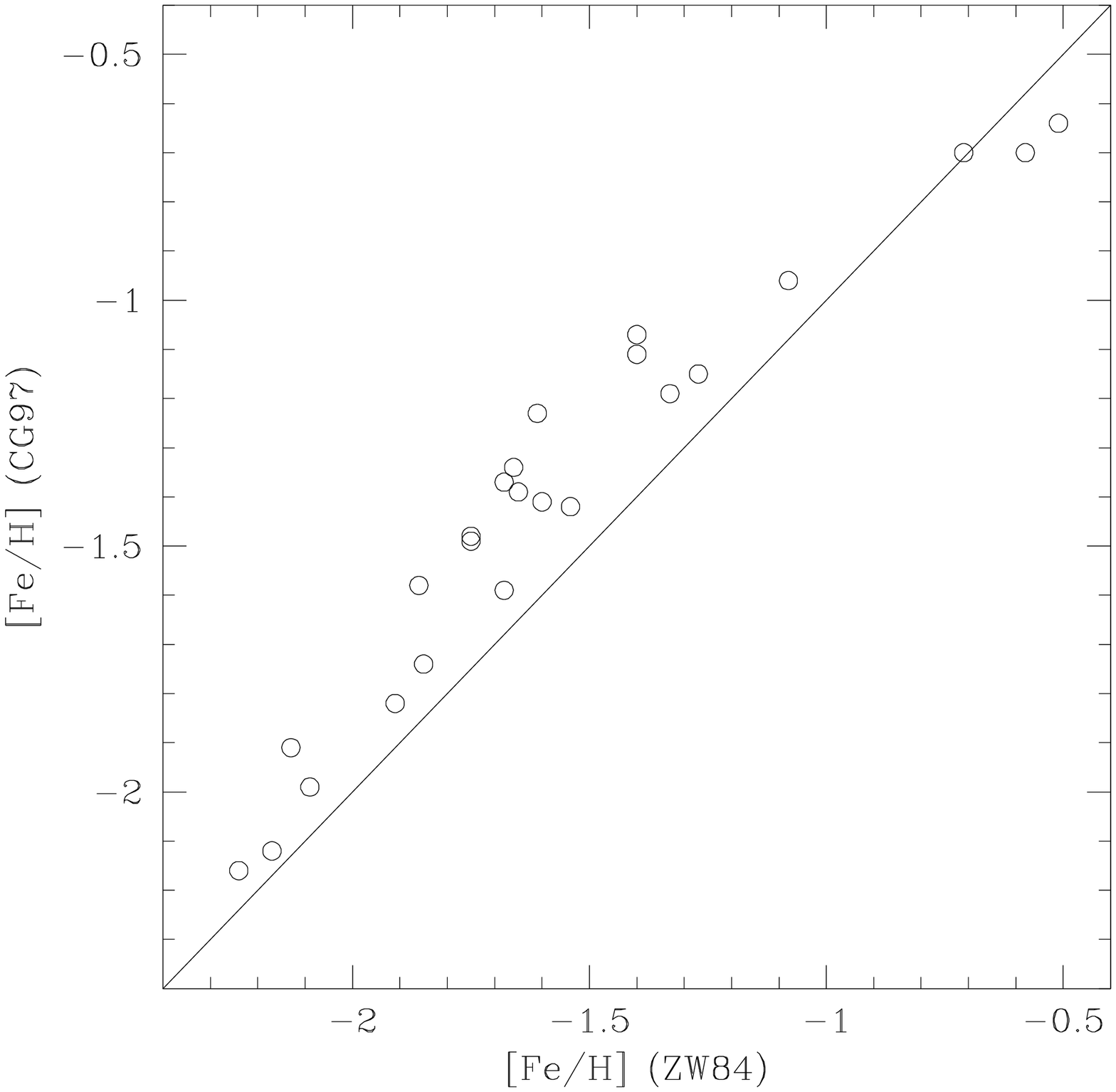}
\caption{Mean metallicities derived by Carretta \& Gratton
(1997) for 24 globular clusters compared against those of 
Zinn \& West (1984) as given in Table 8 of CG97.  The solid 
line indicates where 1:1 agreement would 
lie.\label{figure1}}
\end{figure}

Since 1984, larger samples of giant stars in clusters have 
become available thanks to the construction of larger 
telescopes and improvements in detector technology. In 
addition, availability of more laboratory {\it gf}-values 
plus construction of improved stellar atmosphere models 
have led naturally to a reconsideration of the fundamental 
calibration, the most recent example being that by Carretta 
\& Gratton (1997\nocite{cg97}; hereafter ``CG97''). From 
the available literature, these authors brought together 
CCD-based EWs of more than 160 red giant branch (``RGB'') 
stars in 24 globular clusters to define a new metallicity 
scale.  They adjusted the EWs to a common system and 
re-analyzed all material with a common set of Fe~{\sc i} 
{\it gf}-values, a common set of model atmospheres 
\citep{kurucz92} and a uniform color and effective 
temperature relationship \citep{gratton+96}.  Compared with 
the ZW84\nocite{zw84} calibration, the CG97\nocite{cg97} 
scale led to metallicities which were about 0.1 dex lower 
in [Fe/H] among the most metal-rich ([Fe/H] $>$ --1.0) 
clusters, but which were $\sim$ 0.2 dex higher among many 
clusters of intermediate (--1.0 $<$ [Fe/H] $<$ --1.9) metal 
deficiency.  In Figure~\ref{figure1}, we show the 
relationship between [Fe/H]$_{\rm CG97}$ and 
[Fe/H]$_{\rm ZW84}$ for 24 globular clusters, as given by 
CG97 (their Table 8).

Nevertheless, in the five years since the CG97\nocite{cg97} 
scale was established, some new factors have emerged which 
may affect the abundance analysis of metal-poor giants. 
First, two new scales of color versus effective temperature
(``T$_{\rm eff}$'') have been put forward, one by 
\citet{alonso+99} based on the infra-red flux method 
(``IRFM''; Blackwell {\it et al.}\ 
1990\nocite{blackwell+90}), and another by 
\citet{houdashelt+00} who modelled colors as a function of 
T$_{\rm eff}$ and log~$g$. Second, consideration of the effects 
of non-local thermodynamic equilibrium (``non-LTE'') 
suggests that the assumption of LTE ionization equilibrium 
between Fe~{\sc i} and Fe~{\sc ii} may not be valid, and that 
iron may be overionized in the atmospheres of metal-poor 
giants (Th\'evenin \& Idiart 1999\nocite{ti99}; hereafter 
``TI99''), a conclusion also reached in studies of metal-poor 
RR Lyrae stars \citep{lambert+96,sandstrom+01}. Finally, the 
appearance of accurate Hipparcos \citep{ESA97} parallaxes for 
nearby subdwarfs has led to a small general increase in 
cluster distances, which lowers somewhat the estimates of 
log~$g$ for cluster stars ({\it e.g.}, compare Harris 
1994\nocite{harris94} with the June 22, 1999 revision of 
Harris 1996\nocite{harris96}).

By ``metallicity'', we make the traditional identification of 
equating metallicity with [Fe/H] as derived from the 
analysis of cluster giants.\footnote{
Analyses of small numbers of cluster dwarfs at or near the 
cluster turnoff are slowly becoming available, but not 
enough is known at this time to establish a metallicity 
scale from such stars.}  
In past analyses [Fe/H] has normally 
been based on the lines of Fe~{\sc i} or, in many cases, on 
the mean value of Fe~{\sc i} and Fe~{\sc ii}. However, 
TI99\nocite{ti99}, in a detailed study of LTE models of 
metal-poor stellar atmospheres, concluded that abundance 
analysis of Fe~{\sc i} leads to an underestimate of [Fe/H]. 
TI99\nocite{ti99} contended that, at any given optical depth, 
the populations of the atomic levels of Fe~{\sc i} are 
affected by the outward leakage of UV photons into 
atmospheres made less opaque with decreasing metallicity; 
thus the local kinetic temperature is not in equilibrium with 
the radiation field. The result is that iron suffers from 
overionization. But in such atmospheres, Fe~{\sc ii} is by 
far the dominant species throughout the atmosphere, so that 
the effect of this lack of equilibrium is close to negligible 
for the Fe~{\sc ii} population, but far more serious for 
Fe~{\sc i}. 

The TI99\nocite{ti99} calculation of radiative and collisional 
rates for several hundred Fe~{\sc i} levels led to the 
conclusion that the reduction of [Fe/H] estimated from 
Fe~{\sc i} relative to Fe~{\sc ii} amounts to about 0.1 dex at 
[Fe/H] = --1 and rises to about 0.3 dex at [Fe/H] = --2.5. 
Similar offsets were measured by \citet{lambert+96} in their 
analysis of field RR Lyrae stars.  One would also expect that 
the leakage of UV photons should become larger with lower 
atmospheric densities, {\it i.e.}, surface gravities, at a 
given T$_{\rm eff}$ and with higher T$_{\rm eff}$'s at a given 
luminosity. Just such an effect was found by \citet{ivans+01} 
in their analysis of Fe~{\sc i} and Fe~{\sc ii} abundances 
among first-ascent (RGB) giants of M5 in comparison with 
asymptotic giant branch (``AGB'') giants of comparable 
luminosity but higher T$_{\rm eff}$. They were able to rectify
the abundance of Fe~{\sc ii} on the two branches, but found 
that the traditional LTE abundance analysis led to a 
lowering of the Fe~{\sc i} abundance on the AGB compared 
with the RGB by 0.13 dex, a value essentially in agreement 
with that predicted by TI99\nocite{ti99}.

TI99\nocite{ti99} argued that since Fe~{\sc ii} was by far 
the dominant species of Fe and, in addition, was essentially 
unaffected by departures from LTE, metallicities for globular 
clusters could safely be based on LTE analysis of 
Fe~{\sc ii}, a conclusion supported by \citet{agp01} and
\citet{nissen+02} in their discussions of three-dimensional 
(``3D'') model atmospheres for metal-poor stars. We adopt that 
point of view in what follows, and discuss the techniques that 
need to be applied to obtain reliable values of [Fe/H], based 
on LTE analysis of Fe~{\sc ii} in the spectra of globular 
cluster red giants.  We contend here that, even if the 
TI99\nocite{ti99} conjecture regarding non-LTE effects were
incorrect, it would remain true that, in principle, [Fe/H] is 
more safely derived from the dominant species, Fe~{\sc ii}, 
than from Fe~{\sc i}.

\section{Selection of Stars for Analysis\label{2selection}}

High spectral resolution abundance analyses for giants in 
nearly 30 globular clusters can be found in the literature. 
They differ widely in cluster sample size, some as small as 
one or two stars, others as large as $\sim$35.  The main 
practitioners using modern CCD detectors are comprised of 
four groups:  Gratton and associates (Gratton {\it et al.}\
1986\nocite{gratton+86}, Gratton 1987\nocite{gratton1987}, 
Gratton \& Ortolani 1989\nocite{go89}, CG97\nocite{gc97}); 
the Lick-Texas group (hereafter``LTG''; Sneden {\it et al.}\ 
1994\nocite{sneden+94}, Kraft {\it et al.}\ 
1995\nocite{kraft+95}, Shetrone 1996\nocite{shetrone96}, 
Kraft {\it et al.}\ 1997\nocite{kraft+97}, Sneden 
{\it et al.}\ 1997\nocite{sneden+97}, Ivans {\it et al.}\ 
1999\nocite{ivans+99}, Fulbright 2000\nocite{fulbright00}, 
Shetrone \& Keane 2000\nocite{sk00}, Ivans {\it et al.}
2001\nocite{ivans+01}, but see Kraft 1994\nocite{kraft94} 
for references to earlier papers in the series); Minniti 
{\it et al.}\ 1993\nocite{minniti+93}, 
1996\nocite{minniti+96}; and, most recently, a consortium 
concerned principally with ``metal-rich'' clusters (Cohen 
{\it et al.}\ 1999\nocite{cohen+99}, Carretta {\it et al.}\ 
2001\nocite{carretta+01}, Ram\'irez {\it et al.}\ 
2001\nocite{ramirez+01}).  Contributions more limited in 
scope are those of \citet{mcwilliam+92}, \citet{bw92}, 
\citet{ndc95}, \citet{gw98}, and 
\citet{castilho+00}.\footnote{
More recently, abundances for stars near the main sequence 
in M92 (King {\it et al.}\ 1998\nocite{king+98}), NGC 6397 
(Th\'evenin {\it et al.}\ 2001\nocite{thevenin+01}) and NGC 
6752 (Gratton {\it et al.}\ 2001\nocite{gratton+01}) have 
been obtained. We do not include them in this paper since, 
in the first two references cited, no lines of Fe~{\sc ii}
were measured, and it is also not completely clear at this 
point whether the abundance scale for giants and dwarfs is 
on quite the same system. Further adjustments to the cluster 
metallicity scale may be needed as more analyses of near 
main sequence stars in clusters become available. Special 
remarks concerning a recent study of M71 (Cohen/Ram\'irez
{\it et al.}\ 2001)\nocite{cohen_01,ramirez+01} are found in 
\S~\ref{Arevabunds.g2.m71}.}  
Fortunately, there exist analyses of many giants which are 
common to these investigations.  These stars provide 
information from which transformations in T$_{\rm eff}$, 
log~$g$, {\it etc.}, can be estimated.

CG97\nocite{cg97} adjusted the metallicity determinations of 
the then-available analyses of giants in 24 clusters drawn 
from the groups cited above, using a uniform system for 
determining T$_{\rm eff}$ and log~$g$, a common set of 
{\it gf}-values, and the same set of model stellar 
atmospheres for the program giants and the Sun, those of
\citet{kurucz92}. Owing to the fact that the EW measurements 
of Fe~{\sc ii} lines are either absent or minimal in
several investigations, CG97\nocite{cg97} chose to base 
their metallicity scale on the analysis of Fe~{\sc i} lines. 
However, as a result of the possible departures from LTE for 
Fe~{\sc i} noted in \S~\ref{1intro}, we adopted a 
somewhat different approach. 

We first restricted attention to those clusters in which a 
reasonable number of Fe~{\sc ii} lines, (N(Fe~{\sc ii}) $>$ 
4) were measured in each giant analyzed, and to those 
clusters in which at least three giants were studied.  We 
focussed our attention largely on the work of the LTG 
because the number of stars in a given cluster measured by 
LTG usually proved to be larger than was the case for the 
other groups. In addition, the Fe~{\sc ii} linelists and 
{\it gf}-values chosen varied significantly from group to 
group, with the result that on that account alone, 
transformations between groups often could not be made with 
certainty better than $\sim$0.1 dex. We therefore concluded 
that it would be best to focus attention on 16 clusters, 
uniformly analyzed for Fe~{\sc ii} abundances mostly by 
LTG, ranging in metallicity --2.5 $\lesssim$ 
[Fe/H]$_{\rm II}$ $\lesssim$ --0.7.  We discuss each cluster 
in turn in the Appendix.  However, in light of the 
studies by TI99\nocite{ti99} and \citet{agp01}, we make 
several modifications to the methods employed by LTG, 
including their techniques for estimating T$_{\rm eff}$ and 
log~$g$ and their choices of {\it gf}-values for 
Fe~{\sc ii}.  The present results are therefore essentially 
based on adjustments to LTG's analyses of almost 150 
globular cluster giant stars.

\section{Adjustments to the Input Parameters\label{3adjustments}}

\subsection{Adopted solar Fe abundance; derivation of {\it gf}-values for Fe~{\sc i} and Fe~{\sc ii}\label{3adjustments.solargf}}

     Owing to flux limitations, abundance analyses of globular 
cluster giants are normally confined to the brighter stars 
with T$_{\rm eff}$ $\lesssim$ 4500K; a few investigations have 
dealt with giants as hot as 5000K, often with less than ideal 
signal-to-noise (``S/N''). In such cool stars the flux rolloff 
with decreasing wavelength is rather steep. Thus analysis to 
obtain the Fe~{\sc ii} abundance is usually  confined to 
Fe~{\sc ii} lines with wavelengths $>$ 6000\AA;  LTG derived 
their Fe~{\sc ii} abundances from six lines having wavelengths 
longward of this limit.  In Table~\ref{table1} we list the six 
lines and the log {\it gf}-values adopted by LTG. 

     An extensive investigation of the reliability of 
{\it gf}-values for Fe~{\sc ii} can be found in Lambert 
{\it et al.}\ (1996\nocite{lambert+96}, their Appendix A2). 
The authors concluded that the values derived from theory 
\citep{kurucz93b, nahar95}, experiment \citep{kk87, hk90}, 
and inversion of the solar Fe~{\sc ii} lines for an assumed 
solar iron abundance of log $\epsilon$(Fe$_{\sun}$) = 7.51 
\citep{blackwell+80, biemont+91, hannaford+92} were in 
quite good agreement. However, since experimental values 
are available for relatively few lines (of the six used by 
LTG, only $\lambda$6432.68\AA\ has an 
experimentally-determined {\it gf}-value), they adopted the 
values derived from an assumed solar iron abundance of 7.51. 
We list these also in Table~\ref{table1}, but renormalized 
to a solar log $\epsilon$(Fe) = 7.52, the value adopted by 
LTG.

     Since virtually all of LTG's analyses were based on 
spectra obtained either with the Hamilton \'echelle 
spectrograph \citep{vogt87} operated at the Lick 3.0-m 
telescope or the HIRES spectrograph \citep{vogt+94} at 
the Keck 10.0-m telescope,\footnote{
In the case of M4, most of the spectra were obtained 
with the McDonald Observatory ``2d-coud\'e'' 
spectrograph \citep{tull+95}.  Spectral resolutions for 
the three spectrographs involved range 45,000 $<$ R $<$ 
60,000.} 
we decided to base our log {\it gf}-value 
system on Fulbright's (2000\nocite{fulbright00}) study 
of the solar Fe~{\sc ii} spectrum using daytime sky 
spectra obtained with the Hamilton 
spectrograph\footnote{
EWs from the upgraded (post-1995) Hamilton spectrograph 
are known to be on the system of the Keck~I HIRES 
spectrograph (Shetrone 1996\nocite{shetrone96}, 
Johnson 1999\nocite{johnson99}).  Similarly, by 
consistently comparing the EWs measured in Lick and 
McDonald data for stars observed with both spectrographs, 
LTG have verified the Lick/McDonald EWs are also in 
accord.} 
at the same spectral resolution, grating setting, 
{\it etc.}\ that was used to obtain the spectra of 
globular cluster giants.

     In Table~\ref{table1} we list also the log 
{\it gf}-values adopted by \citet{lambert+96}, the values 
obtained by \citet{blackwell+80}, and those obtained by
\citet{biemont+91}, all adjusted to a solar log 
$\epsilon$(Fe) of 7.52.  The mean difference between 
Blackwell {\it et al.}\nocite{blackwell+80} and those of
\citet{fulbright00} is 0.00, by adoption.  For the three 
lines in common between Lambert 
{\it et al.}\nocite{lambert+96} and 
Fulbright\nocite{fulbright00}, one finds $\Delta$log 
{\it gf} = 0.00 $\pm$ 0.01 ($\sigma$ = 0.02), and 
similarly for the four lines in common between 
Fulbright\nocite{fulbright00} and Biemont 
{\it et al.}\nocite{biemont+91}, $\Delta$log {\it gf} = 
0.00 $\pm$ 0.02 ($\sigma$ = 0.04). However, for the six 
lines in common between LTG and Blackwell 
{\it et al.}\nocite{blackwell+80} (and by inference 
between LTG and Fulbright\nocite{fulbright00}), one 
finds $\Delta$log {\it gf} in the sense  {\it LTG - 
Blackwell {\it et al.}} = --0.06 $\pm$ 0.03 ($\sigma$ = 
0.07).  Thus the log {\it gf}-values adopted by LTG need 
to be increased by 0.06 on average, to conform to the 
system of Blackwell {\it et al.}\nocite{blackwell+80}, 
Lambert {\it et al.}\nocite{lambert+96}, or 
Fulbright\nocite{fulbright00}.

In his analysis, Fulbright\nocite{fulbright00} employed 
a Kurucz model atmosphere \citep{kurucz92, kurucz93a} with 
the adopted input solar parameters of T$_{\rm eff}$ = 5770K, 
log~$g$ = 4.44, log $\epsilon$(Fe) = 7.52, and 
microturbulent velocity (``v$_{t}$'') = 0.84. The 
Kurucz\nocite{kurucz92, kurucz93a} model adopted by 
Fulbright\nocite{fulbright00} was one with convective 
overshooting. However, a Kurucz model {\em without} 
convective overshooting (Castelli, Gratton, \& Kurucz 
1997\nocite{cgk97}) is believed to provide a better 
representation of the solar atmosphere, one that correctly 
predicts the solar spectrum in the ultraviolet shortward of 
$\lambda$3000\AA\ \citep{peterson+02}.  Using such a model, 
we rederived the solar iron abundance from the same input 
parameters and obtained log $\epsilon$(Fe) = 7.51 $\pm$ 0.01 
($\sigma$ = 0.05), based on the EWs of 22 Fe~{\sc ii} 
lines measured by Fulbright.  Thus 
Fulbright's\nocite{fulbright00} assumed log 
{\it gf}-values for Fe~{\sc ii} should be decreased by 
0.01. This suggests that a correction of +0.05 (+0.06 
-- 0.01) is needed to bring the log~{\it gf}-values for 
the six lines measured by LTG into conformity with the 
solar Fe~{\sc ii} abundance.  We note that this small 
adjustment to the log {\it gf}-values of Fe~{\sc ii} 
overall would still maintain close agreement between 
the lab {\it gf}-values and solar-inverted 
{\it gf}-values (see Lambert {\it et al.}\ 
1996\nocite{lambert+96}).

     Although we base our estimates of Fe abundance on 
the analysis of Fe~{\sc ii} lines (denoted by 
[Fe/H]$_{\rm II}$), we make a comparison with abundances 
based on Fe~{\sc i} lines, recognizing that our adopted 
LTE models may give an inadequate representation of the 
true abundance of Fe~{\sc i}.  We therefore need to 
consider also the assignment of log {\it gf}-values to 
the Fe~{\sc i} lines.  Again the situation has been 
thoroughly explored in the paper by \citet{lambert+96}, 
who concluded that the laboratory {\it gf}-values of 
\citet{blackwell+76} and \citet{obrian+91} were in very 
close agreement. Indeed, for 16 lines in common 
considered to have the highest weight, the mean 
difference in log {\it gf} is +0.012 $\pm$ 0.040 in the 
sense of {\it O'Brian\nocite{obrian+91} minus 
Blackwell\nocite{blackwell+76}}. Similar close agreement 
with both O'Brian {\it et al.}\nocite{obrian+91} and 
Blackwell {\it et al.}\nocite{blackwell+76} was found to 
exist with several other investigations ({\it e.g.}, 
Bard {\it et al.}\ 1991\nocite{bard+91}, Bard \& Kock 
1994\nocite{bk94}).  This led Lambert 
{\it et al.}\nocite{lambert+96} to adopt essentially a 
log {\it gf}-value scale combining the work of Blackwell 
{\it et al.}\nocite{blackwell+76} and O'Brian 
{\it et al.}\nocite{obrian+91}

    The LTG Fe~{\sc i} linelist contains nine lines in common 
with those of these two investigations, and from this we find 
a mean difference in log {\it gf}-values of --0.02 $\pm$ 0.07, 
in the sense {\it LTG minus 
(Blackwell\nocite{blackwell+76}, O'Brian\nocite{obrian91})}.  
In addition, we must consider Fulbright's\nocite{fulbright00} 
analysis of the solar Fe~{\sc i} spectrum based on 31 lines. 
Using the Kurucz model without convective overshooting 
\citep{cgk97} as before, employing an interpolation program
kindly provided by Andy McWilliam and subsequently modified,
we derive from Fulbright's EWs a solar Fe~{\sc i} abundance of 
7.54 $\pm$ 0.01 ($\sigma$ = 0.06).  This in turn suggests that 
all Fe~{\sc i}-based values of [Fe/H] (denoted by 
[Fe/H]$_{\rm I}$) derived by LTG should be reduced by 0.04 dex 
(0.02 from LTG versus Blackwell, O'Brian plus 0.02 from 
Fulbright Fe~{\sc i} versus assumed solar of 7.52).

\subsection{Assignment of T$_{\rm eff}$ and estimates of interstellar reddening\label{3adjustments.teffred}}

    Assuming that MARCS models \citep{gustafsson+75} 
adequately represented the atmospheres of metal-poor giants, 
LTG adopted values of T$_{\rm eff}$ based on ``traditional'' LTE 
analysis of the Fe~{\sc i} spectrum.  In this approach, one 
demands that the abundance of Fe~{\sc i} is independent of 
both the excitation potential and strength of the Fe~{\sc i} 
line under analysis.  These demands set both T$_{\rm eff}$ and 
v$_{t}$.  The principal advantage of this approach is that 
it avoids the problem of providing an accurate assessment of 
color excess for any given cluster. 

     However, in our re-analysis of giants considered by 
LTG, we began by abandoning the Fe~{\sc i} excitation plot 
as a means of estimating T$_{\rm eff}$ because of the non-LTE 
precepts discussed by TI99\nocite{ti99}. Instead, we 
estimated T$_{\rm eff}$ for each giant from its observed 
reddening-corrected value of $V$--$K$ \citep{cohen+78, 
frogel+81, frogel+83} when available, otherwise $B$--$V$, 
applying the values of color excess given in the 
June 22, 1999 revision of the Harris (1996)\nocite{harris96} 
compilation. The color versus T$_{\rm eff}$ scale adopted is 
that of Alonso {\it et al.}\ (1999\nocite{alonso+99}, their 
Table 6),\footnote{
We have since verified that the corrected version of the 
formula given by \citet{alonso+01} for the caption to Table 
2 of \citet{alonso+99} corresponds to the values given in 
Table 6 of \citet{alonso+99}.}
based on the IRFM \citep{blackwell+90}.  However, use of 
colors requires one to consider the effect of errors in 
$E(B$--$V)$ on the derived value of T$_{\rm eff}$.  This, in 
turn, depends on the color and magnitude of the particular 
star being analyzed.  For example, in a cluster with [Fe/H] = 
--2.0, a giant having $B$--$V$ = 1.22 and corresponding 
T$_{\rm eff}$ = 4250K for $E(B$--$V)$ = 0.00 would have its 
T$_{\rm eff}$ increased by 60K to 4310K if $E(B$--$V)$ = 0.05.  
But for a fainter giant in the same cluster having $B$--$V$ = 
0.71 and corresponding T$_{\rm eff}$ = 5000K if unreddened, an 
increase in $E(B$--$V)$ of 0.05 corresponds to an increase in 
T$_{\rm eff}$ to 5150K, a 150K change. However, the corresponding 
change in the abundance of Fe~{\sc ii}, {\it viz}, a decrease 
of $\sim$0.02 dex (see Kraft {\it et al.}\ 
1997\nocite{kraft+97}, their Table 5), is essentially the 
same, independent of T$_{\rm eff}$.  On the other hand, the change 
in Fe~{\sc i} is much larger: an increase of $\sim$ +0.12 dex.  
In the case of more metal-rich cluster giants, the effect on 
Fe~{\sc ii} is larger (see, {\it e.g.}, Shetrone 
1996\nocite{shetrone96}, his Table 6 for M13, or Ivans 
{\it et al.}\ 2001\nocite{ivans+01}, their Table 3 for M5), 
amounting to --0.08 dex for Fe~{\sc ii} and +0.07 dex for 
Fe~{\sc i}, again for an increase in $E(B$--$V)$ = 0.05.  
For this reason, we focus most of our attention on calibration 
clusters having relatively well-known small color excesses, 
{\it i.e.}, $E(B$--$V)$ $\lesssim$ 0.10.  However, we also 
include such exceptions as M4 and M10, which will be discussed 
in \S~\ref{Arevabunds.g2}.

However, we found that despite the non-LTE uncertainties 
associated with the employment of the Fe~{\sc i} spectrum as a 
means of estimating T$_{\rm eff}$, the values of T$_{\rm eff}$ so 
obtained in most cases agree surprisingly well with those 
based on colors and use of the \citet{alonso+99} calibration. 
A comparison is made in Table~\ref{table2} for a number of key 
clusters, and it will be seen that the difference in T$_{\rm eff}$ 
is always less than $\sim$40K among clusters with small and 
well-determined color excesses.  The difference of --39K for M13 
is the largest among such clusters.  The effect of a 40K change 
in T$_{\rm eff}$ leads to a 0.03 dex change in the Fe~{\sc ii} 
abundance of M3, M13, M5 and M10, but only about 0.01 dex change 
at the lower metallicities of M92 and M15. Thus whether we had 
chosen to use the Fe~{\sc i} excitation plot or the $V$--$K$ (or 
$B$--$V$) colors to obtain T$_{\rm eff}$ has only a small bearing on 
the derived abundances of Fe~{\sc ii} in the cases of M5, M3, 
M13, M92 or M15.

      The mean value of $\Delta$T$_{\rm eff}$ for the seven clusters 
of Table~\ref{table2} having small and well-determined color 
excesses and for which each $\Delta$T$_{\rm eff}$ $<$ $|$40K$|$ is 
+2K $\pm$ 20K ($\sigma$ = 53K). Such close agreement between the 
T$_{\rm eff}$'s based on the Fe~{\sc i} excitation plot and the 
\citet{alonso+99} color calibration for little-reddened 
clusters encourages us to adopt the former in those cases in 
which interstellar reddening estimates are large or 
controversial, as for example, clusters such as M10 and M71. 
We discuss these in more detail in \S~\ref{Arevabunds.g1} and 
\ref{Arevabunds.g2}.

      The Fe~{\sc i} excitation plot therefore provides a new 
means of estimating color excesses for clusters: by identifying 
the values of T$_{\rm eff}$ determined from the excitation plot on 
the one hand and the \citet{alonso+99} calibration on the other, 
we can estimate the value of $E(B$--$V)$ that brings the T$_{\rm eff}$
estimates into agreement. The reliability of $E(B$--$V)$ obtained 
from this application of the Fe~{\sc i} excitation plot can be 
tested by comparing the results with values of $E(B$--$V)$ given 
by the \citet{schlegel+98} dust maps. These are listed also in 
Table~\ref{table2}, and are found to be in remarkably good 
agreement with the excitation-plot values for two clusters with 
fairly large and/or uncertain reddenings, M71 and NGC 7006, 
where $\Delta$$E(B$--$V)$ $\leq$ 0.02 mag for both. The agreement 
in the case of little-reddened clusters is also excellent.  Only 
for M10 is the situation somewhat disappointing: here the 
disagreement in $E(B$--$V)$ is 0.05 mag. The only other cluster for 
which there is a large difference is M4: but this is 
understandable since \citet{schlegel+98} assume a ratio of total 
to selective absorption of $R$ = 3.1, whereas it is known that 
this ratio is anomalously large in the case of M4 (see Ivans
{\it et al.}\ 1999\nocite{ivans+99} and references therein).

      In the preceding, we adopted values of T$_{\rm eff}$ based on 
the \citet{alonso+99} color versus T$_{\rm eff}$ scale, which makes 
use of the IRFM.  On the other hand, the recent 
\citet{houdashelt+00} scale, based on modelling stellar colors, 
yields values of T$_{\rm eff}$ that are systematically larger. In 
the case of M5, \citet{ivans+01} found a mean offset of +43K in 
the sense {\it T$_{\rm eff}$(Houdashelt) minus T$_{\rm eff}$(Alonso)}, 
but the offset proved to be T$_{\rm eff}$-dependent, varying from 
$\sim$ zero at T$_{\rm eff}$ of 4700K to $\sim$ +60K at a T$_{\rm eff}$ 
of 3900K.  We generally find offsets averaging near 100K for M3 
and M13 giants but slightly smaller offsets near 75K for the more 
metal-poor clusters M15 and M92. We discuss more fully the 
effect of adopting the \citet{houdashelt+00} T$_{\rm eff}$ scale 
on the Fe~{\sc ii} (and Fe~{\sc i}) abundances in 
\S~\ref{6adopthoud}.

\subsection{Determination of log~$g$: cluster moduli and stellar masses\label{3adjustments.gmass}}

     Fundamental considerations give rise to the following 
equation for the surface gravity of a star:
\begin{equation}
log \frac{g_{\star}}{g_{\sun}} = log \frac{M_{\star}}{M_{\sun}} - 4 log \frac{5770}{T_{eff_{\star}}} - 0.4(4.72 - M_{bol}).
\end{equation}

\noindent Present ideas of stellar evolution coupled with 
current estimates of the ages of the oldest, metal-poor 
stars closely constrain the value of the mass 
(``$M_{\star}$'') because the turnoff mass is a very slow 
function of age.  Thus for an age of 13 Gyr and a hydrogen 
content of 0.75, $M_{\star}$ = 0.80~$M_{\sun}$ 
\citep{vandenberg00}. If the age is allowed to decrease by 
2 Gyr to take into account such clusters as Rup 106, 
$M_{\star}$ increases to 0.84~$M_{\sun}$; an age increase 
to 15 Gyr decreases the mass to 0.775~$M_{\sun}$. Thus the 
expected value of log $M_{\star}$/$M_{\sun}$ is --0.097 
with a range of only $\pm$ 0.02. Thus once one chooses a 
value of T$_{\rm eff}$, the appropriate value of log~$g$ is 
constrained entirely by the choice of cluster true 
(reddening-freed) distance modulus and the value of the 
bolometric correction. 

      LTG obtained estimates of log~$g$ based solely on the 
analysis of the EWs of the Fe~{\sc i} and Fe~{\sc ii} lines, 
with the requirement that the total iron abundance derived 
from Fe~{\sc i} essentially equals that derived from 
Fe~{\sc ii}.  But, in light of the non-LTE calculations of 
TI99\nocite{ti99}, this may not, in principle, be correct.  
We therefore take up consideration of the reliability of 
cluster distance modulus determinations.

     Recent estimates, {\it e.g.}, \citet{gratton+97} and 
\citet{reid97, reid98}, have been based on fitting 
color-magnitude diagrams of cluster main sequences to colors 
and absolute magnitudes of subdwarfs observed by Hipparcos
having metallicities similar to those of the clusters.  An 
update of the procedures employed in the first of these 
papers is found in \citet{carretta+00}, which contains a 
thorough discussion of the problems involved. In all papers, 
the necessary \citet{lutz-kelker73} corrections were applied 
and found to be very small. The primary lists of stars 
consisted of Hipparcos subdwarfs having relative parallax 
errors $\sigma\pi$/$\pi$ $<$ 0.12, although each paper 
also lists stars having larger parallax errors. The authors 
paid close attention to the influence of interstellar 
reddening, and carefully weeded the lists for the presence 
of binaries.  This is especially the case in the 
\citet{gratton+97} investigations. Attention was also paid 
to the fact that subdwarfs having M$_{V}$ $<$ +5.5 almost 
certainly undergo some evolutionary departure from the 
zero-age main sequence in the age interval 11-15 Gyr. Thus 
main sequence cluster fitting was based mostly on stars 
with M$_{V}$ $>$ +6.0.

    Nevertheless uncertainties remain, especially in the 
treatment of the most metal-poor clusters such as M92, 
since there is a paucity of Hipparcos subdwarfs satisfying 
the restrictions that [Fe/H] $<$ --2.0, M$_{V}$ $>$ +6.0, 
and $\sigma\pi$/$\pi$ $<$ 0.12. For example, in Reid's 
list of subdwarfs having $\sigma\pi$/$\pi$ $<$ 0.12, 
there is only one star with M$_{V}$ $>$ +6 and [Fe/H] $<$ 
--2.0 (BD+66~268), an astrometric binary. Similarly, in 
the list of \citet{carretta+00}, which contains 56 
subdwarfs satisfying the parallax error limit cited above, 
the most metal-poor single star (M$_{V}$ $>$ +6.0) has a 
metallicity [Fe/H] = --1.92 (G246-38) on the 
CG97\nocite{cg97} scale in which the metallicity of M92 is 
[Fe/H] = --2.15.\footnote{
The star HD 46663 satisfies the parallax error limit and 
has M$_{V}$ $>$ +6.0, but its color is much too red for 
its M$_{V}$. It too is an astrometric binary.}  
On the 
ZW84\nocite{zw84} scale the metallicity of M92 is --2.24, 
and later in this paper we derive --2.38. 

Although it is generally conceded that the color shifts 
necessary to {\em correct} the most metal-poor subdwarf 
colors toward lower metallicity are small, they are not 
negligible and do require an extrapolation.  For example, 
subdwarfs with M$_{V}$ $\sim$ +6.5 and having [Fe/H] =
--2.3 are 0.03 mag bluer in $B$--$V$ than those having 
[Fe/H] = --2.0 (see {\it e.g.} VandenBerg {\it et al.}\ 
2000\nocite{vandenberg00}) and this corresponds to a 
shift toward a fainter absolute magnitude of 0.15 mag.  
Indeed, Gratton {\it et al.}\ fit the color-magnitude 
array of M92 at an extrapolated fixed position: $B$--$V$ = 
0.55 and M$_{V}$ = +6.0, which gives a good fit to the 
``corrected'' position of only one sufficiently faint 
subdwarf, the well-known HD~25329, a star with 
overabundances of N and neutron-capture nucleosynthesis 
process species ({\it e.g.}, Ba, La, Ce, Nd, and Sm; 
Beveridge \& Sneden 1994\nocite{bs94}).  Although these 
abundance peculiarities probably have no effect on the 
luminosity of this star (it is no doubt running on the 
p-p chain), its color could be somewhat affected by 
excessively strong bands of CN at $\lambda\lambda$3883, 
4215\AA. In any case, its metallicity of --1.84 
\citep{bs94} or --1.69 \citep{cg97} is rather too high 
to provide a good match for M92.

     These problems associated especially with the 
paucity of Hipparcos-based low metallicity subdwarfs 
were fully recognized by the authors of the papers 
cited above and come as no surprise. Accordingly, we 
decided to re-derive the true distance moduli, 
(m--M)$_{0}$, for five key low-metallicity clusters, in 
part because some of these key clusters were not 
considered by \citet{carretta+00}.  We used only those 
subdwarfs with M$_{V}$ $>$ +6.0 having metallicities that 
closely matched the metallicity of the cluster under 
consideration, but relaxed the restrictions on the 
Hipparcos-based parallax errors. 

All stars were drawn from the tables presented in the 
aforementioned papers.  Thus the reddening and 
absorption-corrected color-magnitude arrays for M92 and M15 
were adjusted to fit the M$_{V}$'s of the four stars in 
Reid's list having the lowest values of [Fe/H]: --2.10 
(BD+66~268), --1.91 (BD+59~2407), --2.26 (G254-24), and 
--2.68 (BD+38~4955), despite the fact that one of these 
stars is a known single-line binary and the other is a 
suspected binary. These last two were given low weight in 
the procedure of fitting the color-magnitude arrays of M92 
and M15 ``by eye'' to the M$_{V}$ and $B$--$V$ values of 
these stars. Our estimates of (m--M)$_{0}$ for M92, M15, 
M13, M3 and M5 are given in Table~\ref{table3} and compared 
with prior estimates. For M13 and M92, our estimates agree 
with those of \citet{vandenberg00} to within $\sim$0.1 mag.

     The agreement between the present moduli and the 
estimates of Reid and Gratton {\it et al.}\ is generally 
good, with a modulus range difference $\lesssim$ 0.2 mag. A 
comparison with the apparent moduli derived from the 
June 22, 1999 revision of the \citet{harris96} compilation 
is also good. For the five key clusters under consideration, 
we find $\Delta$(m--M)$_{V}$ = +0.07 $\pm$ 0.04 ($\sigma$ = 
0.11) in the sense {\it this study minus Harris}. On the 
other hand, for nine little-reddened clusters studied by 
Carretta {\it et al.}\ one finds $\Delta$(m--M)$_{V}$ = 
+0.16 $\pm$ 0.03 ($\sigma$ = 0.09) in the sense {\it Carretta 
{\it et al.}\ minus Harris}. The Carretta {\it et al.}\ 
moduli run about 0.1 mag larger than the present values. 
Since in this paper we deal with a number of clusters not 
considered by Carretta {\it et al.}, and since the Harris 
compilation contains essentially all the known clusters, 
we decided to adopt the Harris values increased by 0.07 
mag. We note that decreases of order 0.1 dex in moduli do 
not seriously affect the derived Fe~{\sc ii} abundances, 
since they generate an increase of 0.04 dex in log~$g$ and 
a corresponding increase in [Fe/H]$_{\rm II}$ of only 0.01.

     The possibility of systematic errors in the bolometric 
corrections also must be considered. LTG adopted the values 
given in tables by \citet{worthey94}, based on the 
\citet{kurucz92, kurucz93a} grid of model atmospheres.  The 
more recent determinations of \citet{houdashelt+00} are 
slightly smaller than those based on Kurucz, by amounts 
varying from 0.04 to 0.09 mag in the range of relevant 
T$_{\rm eff}$'s, log~$g$'s and metallicities. This amounts at 
most to an increase in log~$g$ of 0.04 dex, and corresponding 
increase in [Fe/H]$_{\rm II}$ of $\sim$0.01. This being 
negligible, we simply continued to employ the values in the 
tables from Worthey.

\subsection{Abundance adjustments: Strategy\label{3adjustments.strategy}}

     We divide our clusters into three groups: (1) key 
fundamental clusters having small color excesses ($E(B$--$V)$ 
$\lesssim$ 0.10), in which there is LTG analysis of 
Fe~{\sc i} and Fe~{\sc ii} lines of at least six giants, and 
for which T$_{\rm eff}$ and log~$g$ can be estimated by appeals 
to colors and absolute magnitudes; (2) secondary clusters 
again having LTG analysis of at least six stars but for 
which $E(B$--$V)$ $>$ 0.10 and for which we therefore substitute 
T$_{\rm eff}$ values derived from the Fe~{\sc i} excitation plot, 
in accordance with the results tabulated in Table~\ref{table2}
and discussed in \S~\ref{3adjustments.teffred}; and (3) 
tertiary clusters derived from selected investigations 
outside the LTG. We set here a somewhat arbitrary
requirement that the analysis should be based on a study of 
at least three giant stars per cluster.

     Since we are concerned principally with lines of 
Fe~{\sc ii}, we keep the same v$_{t}$'s as those derived 
in the original papers, even though these were derived 
from the Fe~{\sc i} line spectrum. In all clusters, 
independent of metallicity, the EWs of Fe~{\sc ii} lines 
rarely exceed 50 m\AA, and thus lie near or on the linear 
portion of the curve of growth and are thus little 
affected by variations of order $\pm$ 0.2 dex in v$_{t}$. 
Even in the most metal-rich cluster considered by LTG 
(M71; Sneden {\it et al.}\ 1994), out of 10 giants in which 
45 Fe~{\sc ii} EWs were measured, only nine lines had EWs 
exceeding 60 m\AA.  Lines of Fe~{\sc i}, being mostly 
stronger than those of Fe~{\sc ii}, are more seriously 
affected, but Fe~{\sc i} is not fundamental to the abundance 
scale discussed here.

     Having established the values of T$_{\rm eff}$ and log~$g$ 
from colors and absolute magnitudes, we then ``correct'' the 
Fe~{\sc ii} (and Fe~{\sc i}) abundances for stars in Group 1 
in accordance with the tables of adjustments in the LTG 
papers cited in the Appendix.  For Group 2 stars, the 
adjustments mostly involve log~$g$, because of recent 
updates in cluster distance moduli.  We take up each cluster 
in turn, discuss the modifications, and extend the discussion 
in more detail for those clusters in Group 3.

     We adopt MARCS models for the analysis of giants, 
following the analysis procedure of LTG. We also employed 
Kurucz models distributed over the abundance range of 
interest in the analysis of a few representative cluster 
giants and determined the offsets in log 
$\epsilon$(Fe~{\sc i}) and log $\epsilon$(Fe~{\sc ii}) 
between Kurucz and MARCS models.  The offsets are applied to 
provide an abundance scale based on Kurucz as well as MARCS 
models. We also discuss in \S~\ref{6adopthoud} the offsets 
that need to be applied if one employs the Houdashelt 
{\it et al.}\ values of T$_{\rm eff}$ in place of the Alonso 
{\it et al.}\ values.  Finally we discuss the relationship 
between the values of [Fe/H]$_{\rm II}$ derived here and the 
low resolution reddening-free spectroscopic parameter 
$<$$W'$$>$ introduced by \citet{rutledge+97a, rutledge+97b}, 
so that the derived metallicity scale can be applied to a 
large population of galactic globular clusters.

\section{Revised Cluster Abundances Based on Fe~{\sc ii}\label{4revabunds}}

     We take up each of the Groups 1 to 3 in 
Appendices~\ref{Arevabunds.g1}, \ref{Arevabunds.g2}, and 
\ref{Arevabunds.g3}, respectively.  Within each Group, we discuss 
each cluster in detail in order of decreasing metallicity.  We 
make the {\it ab initio} assumptions that MARCS models and the 
Alonso {\it et al.}\ T$_{\rm eff}$ scale apply, and discuss further 
the necessary transformations to put the clusters on Kurucz models 
in \S~\ref{5transform} and the Houdashelt {\it et al.}\ 
T$_{\rm eff}$ scale in \S~\ref{6adopthoud}. Here we give a catalog of 
the clusters considered and their location in the Appendices.

\subsection{Group 1 Clusters: T$_{\rm eff}$ and log~$g$ based on colors 
and absolute magnitudes}
\begin{tabbing}
123\=456\=789\=\kill
\> Appendix~\ref{Arevabunds.g1.m5} M5 (NGC 5904); $E(B$--$V)$ = 0.03, (m--M)$_{0}$ = 14.42\\
\> Appendix~\ref{Arevabunds.g1.n362} NGC 362; $E(B$--$V)$ = 0.05, (m--M)$_{0}$ = 14.70\\
\> Appendix~\ref{Arevabunds.g1.n288} NGC 288; $E(B$--$V)$ = 0.03, (m--M)$_{0}$ = 14.66\\
\> Appendix~\ref{Arevabunds.g1.m3} M3 (NGC 5272); $E(B$--$V)$ = 0.01, (m--M)$_{0}$ = 15.02\\
\> Appendix~\ref{Arevabunds.g1.m13} M13 (NGC 6205); $E(B$--$V)$ = 0.02, (m--M)$_{0}$ = 14.42\\
\> Appendix~\ref{Arevabunds.g1.m92} M92 (NGC 6341); $E(B$--$V)$ = 0.02, (m--M)$_{0}$ = 14.75\\
\> Appendix~\ref{Arevabunds.g1.m15} M15 (NGC 7078); $E(B$--$V)$ = 0.10, (m--M)$_{0}$ = 15.25\\
\end{tabbing}

\subsection{Appendix~\ref{Arevabunds.g2} Group II Clusters: Large and/or Uncertain Reddening}
\begin{tabbing}
123\=456\=789\=\kill
\> Appendix~\ref{Arevabunds.g2.m71} M71 (NGC 6838); $E(B$--$V)$ = 0.32, (m--M)$_{0}$ = 12.83\\
\> Appendix~\ref{Arevabunds.g2.m4} M4 (NGC 6171); $E(B$--$V)$ = 0.33$_{variable}$, (m--M)$_{0}$ = 11.61\\
\> Appendix~\ref{Arevabunds.g2.m10} M10 (NGC 6254); $E(B$--$V)$ = 0.24, (m--M)$_{0}$ = 13.41\\
\end{tabbing}

\subsection{Appendix~\ref{Arevabunds.g3} Other Clusters; Other Sources}
\begin{tabbing}
123\=456\=789\=\kill
\> Appendix~\ref{Arevabunds.g3.probs} Some Transformation Problems\\
\> Appendix~\ref{Arevabunds.g3.47tuc} 47 Tuc (NGC 104); $E(B$--$V)$ = 0.04, (m--M)$_{0}$ = 13.32\\
\> Appendix~\ref{Arevabunds.g3.n7006} NGC 7006; $E(B$--$V)$ = 0.10, (m--M)$_{0}$ = 18.00\\
\> Appendix~\ref{Arevabunds.g3.n3201} NGC 3201; $<$$E(B$--$V)$$>$ = 0.25, (m--M)$_{0}$ = 13.61\\
\> Appendix~\ref{Arevabunds.g3.n6752} NGC 6752; $E(B$--$V)$ = 0.04, (m--M)$_{0}$ = 13.07\\
\> Appendix~\ref{Arevabunds.g3.n2298} NGC 2298; $E(B$--$V)$ = 0.16, (m--M)$_{0}$ = 15.17\\
\> Appendix~\ref{Arevabunds.g3.n6397} NGC 6397; $E(B$--$V)$ = 0.24, (m--M)$_{0}$ = 11.62\\
\end{tabbing}

\section{Transformation to Kurucz Models\label{5transform}}

     In the preceding section and Appendices, we assumed the 
correctness of MARCS models in analyzing the spectra of globular 
cluster giants.  Kurucz model atmospheres are also available, 
and they are of two types: with or without convective overshooting. 
We have already noted the contention of \citet{peterson+02} that 
the solar spectrum shortward of $\lambda$3000\AA, as well as the 
spectra of several metal-poor dwarfs, can best be fit using 
Kurucz models without convective overshooting. As we have 
seen, many cluster giant star analyses, those, for example, 
by CG97 and by Minniti {\it et al.}, have used Kurucz models 
with convective overshooting.  The question is this: which of 
the three possible model atmospheres is best when one wishes 
to analyze the spectrum of a globular cluster giant?

     In the case of the Sun and the Peterson dwarfs, the 
Kurucz models without convective overshooting are clearly 
favored. But there is nothing to guide us in the choice of 
correct models for globular cluster giants. It does not 
follow that, just because Kurucz models without overshooting 
provide satisfactory representations of the UV spectra of 
dwarfs, these models are the ones to use in the case of 
globular cluster giants. We take the position here that until 
further evidence is forthcoming, the three types of models
have an equal claim to validity. Consequently we examine here 
what happens to the Fe~{\sc i} and (especially) Fe~{\sc ii} 
abundances when we employ Kurucz models with and without 
convective overshooting.

     To deal with this question, we decided not to try to run 
all 149 giants of this study through two sets of Kurucz 
models, but rather to run representative cluster giants, ones 
already analyzed via MARCS models, through Kurucz models 
using exactly the same choices of T$_{\rm eff}$, log~$g$, and 
input abundances. To cover the range in metallicity studied
here, we chose giants in M71 ([Fe/H]$_{\rm II}$ = --0.78), M3 
([Fe/H]$_{\rm II}$ = --1.45), and M92 ([Fe/H]$_{\rm II}$ = 
--2.38) at typical T$_{\rm eff}$'s in the 4200 to 4400K range and 
typical log~$g$'s in the 0.5 to 1.0 range. 

     The results are summarized as follows. The shifts appear 
to be fairly insensitive to changes in T$_{\rm eff}$ (and 
corresponding log~$g$) over an interval of $\sim$400K at a 
fixed metallicity.  At all metallicities, there is virtually
no difference in log $\epsilon$(Fe~{\sc ii}) whether one 
uses Kurucz models with, or without, convective overshooting. 
However, compared with MARCS models, the Kurucz models always 
lead to higher values of log $\epsilon$(Fe~{\sc ii}): in M71 
by 0.06, in M3 by 0.08 (on the average), and in M92 by 0.06 
again. 

The situation for Fe~{\sc i} is more complex. At the 
highest (--0.78) and intermediate (--1.45) metallicities, 
there is virtually no change in log $\epsilon$(Fe~{\sc i}) 
between the two kinds of Kurucz models, but at the lowest 
metallicity (--2.38), the models without overshooting run 
0.08 dex smaller than the models with overshooting, and are 
quite close to the results from MARCS models. The MARCS models 
yield log $\epsilon$(Fe~{\sc i}) values close to those of the 
Kurucz models.  On the other hand, at the high metallicity 
end, MARCS models are more metal-rich than Kurucz models by 
$\sim$0.07, whereas at the lowest metallicity end of the 
scale, they are more metal-poor than the Kurucz models with
overshooting by about the same amount, although they are in 
substantial agreement with the Kurucz models without 
overshooting. 

Since our focus is on Fe~{\sc ii}, the conclusion is fairly 
straightforward: Kurucz models with or without convective 
overshooting yield Fe~{\sc ii} metallicities 0.06 to 0.08 
dex higher than MARCS models, independent of metallicity in 
the regime --2.4 $<$ [Fe/H] $<$ --0.7.

     We summarize the Fe~{\sc i} and Fe~{\sc ii} abundance 
estimates of the clusters in Groups 1, 2, and 3 in 
Table~\ref{table4}. In columns 2 and 3, we list the values 
of [Fe/H]$_{\rm II}$ and corresponding $\sigma$, respectively, 
and in columns 4 and 5, the same for [Fe/H]$_{\rm I}$, as 
derived (see text) from MARCS models. In columns 6 and 7 
are the abundances of Fe~{\sc ii} and Fe~{\sc i} based on 
Kurucz models with convective overshooting, and columns 8 
and 9 the corresponding values based on Kurucz models 
without convective overshooting.  Owing to the method 
employed here, the $\sigma$'s for columns 6 to 9 are, by 
construction, the same as those listed in columns 3 and 5.

\begin{figure}
\plotone{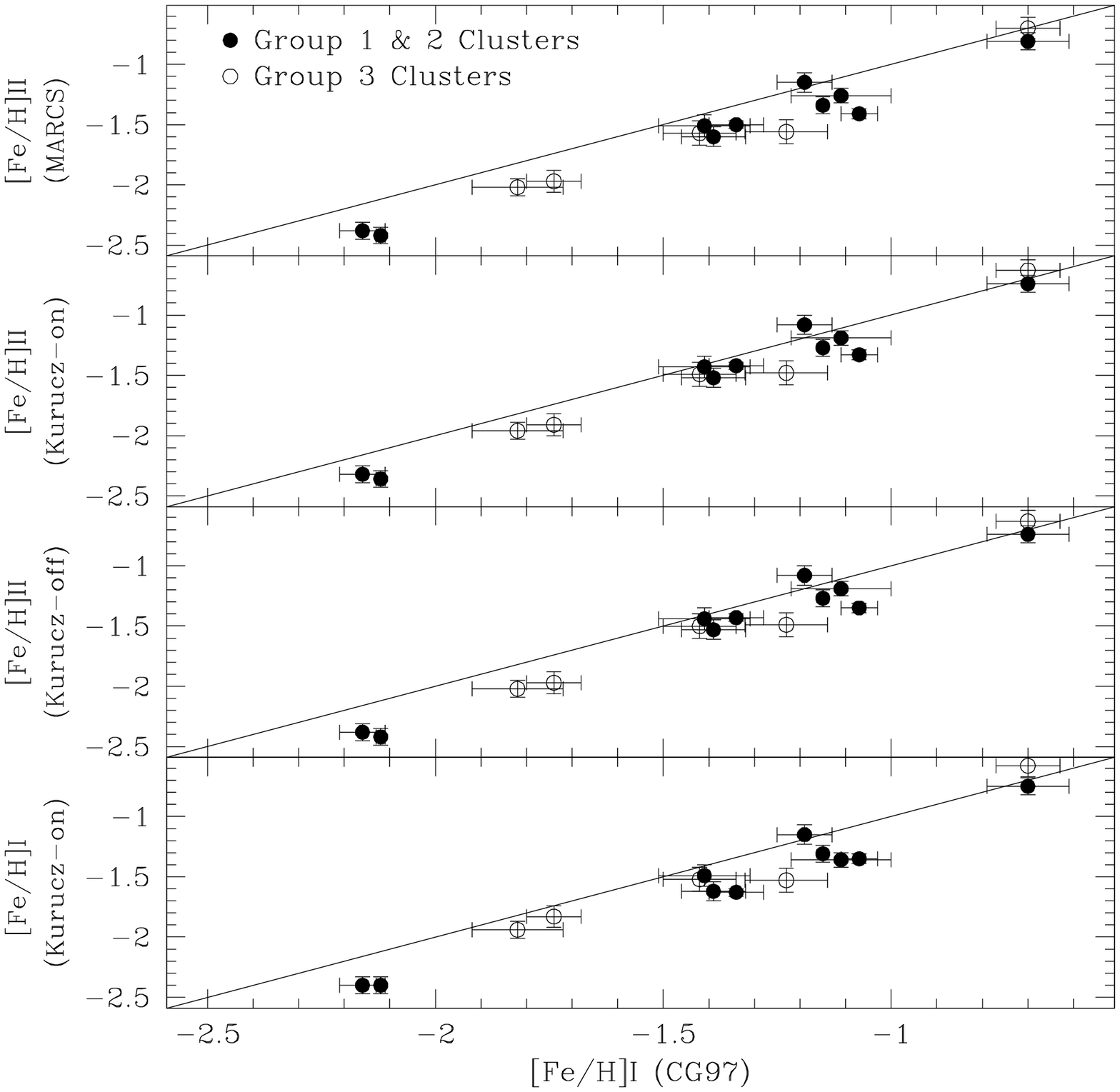}
\caption{The top panel displays our results from 
Table~\ref{table4} for [Fe/H]$_{\rm II}$ employing MARCS 
models as a function of the [Fe/H] scale derived by 
\citet{cg97}; the second panel from the top displays the same 
information for Kurucz models with overshooting; the third 
panel displays the same information for Kurucz models without 
overshooting; and the bottom panel displays our results for 
[Fe/H]$_{\rm I}$ employing MARCS models.  In solid symbols 
are those clusters designated Groups 1 and 2; in open symbols 
are those designated as belonging to Group 3 (see 
\S~\ref{3adjustments.strategy} and \S~\ref{5transform} for 
details).  The solid line in each panel indicates where 1:1 
agreement would lie.\label{figure2}}
\end{figure}

     In Figure~\ref{figure2} we plot as abscissa [Fe/H]$_{\rm I}$, 
as derived by CG97 from analysis of Fe~{\sc i} using Kurucz models 
with convective overshooting, against [Fe/H]-values derived here 
on the basis of three different assumptions: (1) [Fe/H]$_{\rm II}$ 
from MARCS models (Table~\ref{table4}, column 2), (2) 
[Fe/H]$_{\rm II}$ from Kurucz models with convective overshooting 
(Table~\ref{table4}, column 6) and (3) [Fe/H]$_{\rm II}$ from Kurucz 
models without convective overshooting (Table~\ref{table4}, column 
7). We also compare in the bottom panel [Fe/H]$_{\rm I}$(CG97) with 
[Fe/H]$_{\rm I}$ derived here.  

     In all cases the main effect seems to be that the present
abundances are smaller than those of CG97, and this is true even 
when comparing the values of [Fe/H]$_{\rm I}$ for a common Kurucz 
model, the one with convective overshooting (bottom panel).  The 
offset is between zero at the highest metallicity, and 0.2 dex 
at the lowest. About half of the offset is probably attributable 
to the difference in adopted values of T$_{\rm eff}$: CG97's values 
of T$_{\rm eff}$ run generally 50 to 100K hotter than the values 
adopted here, but are nearly zero in the case of M71. This would
account for 0.05 to 0.12 dex of the offset for the more 
metal-poor clusters.  In addition, the lower values of log~$g$ 
adopted here (in comparison with those adopted by CG97) would 
also lead to slightly smaller values of [Fe/H]$_{\rm I}$.  On the 
other hand, lower surface gravities have more serious effect in 
lowering abundances of Fe~{\sc ii}, and this may well account for 
the still larger offset seen in the top panel of 
Figure~\ref{figure2}. 

{\it The main point to be made here is simply that different 
choices of T$_{\rm eff}$-scales, values of log~$g$, and models can 
easily introduce systematic uncertainties in the globular cluster 
abundance scale, and these systematic effects can be as large as 
0.1 and up to 0.3 dex in the derived value of [Fe/H].}

\begin{figure}
\plotone{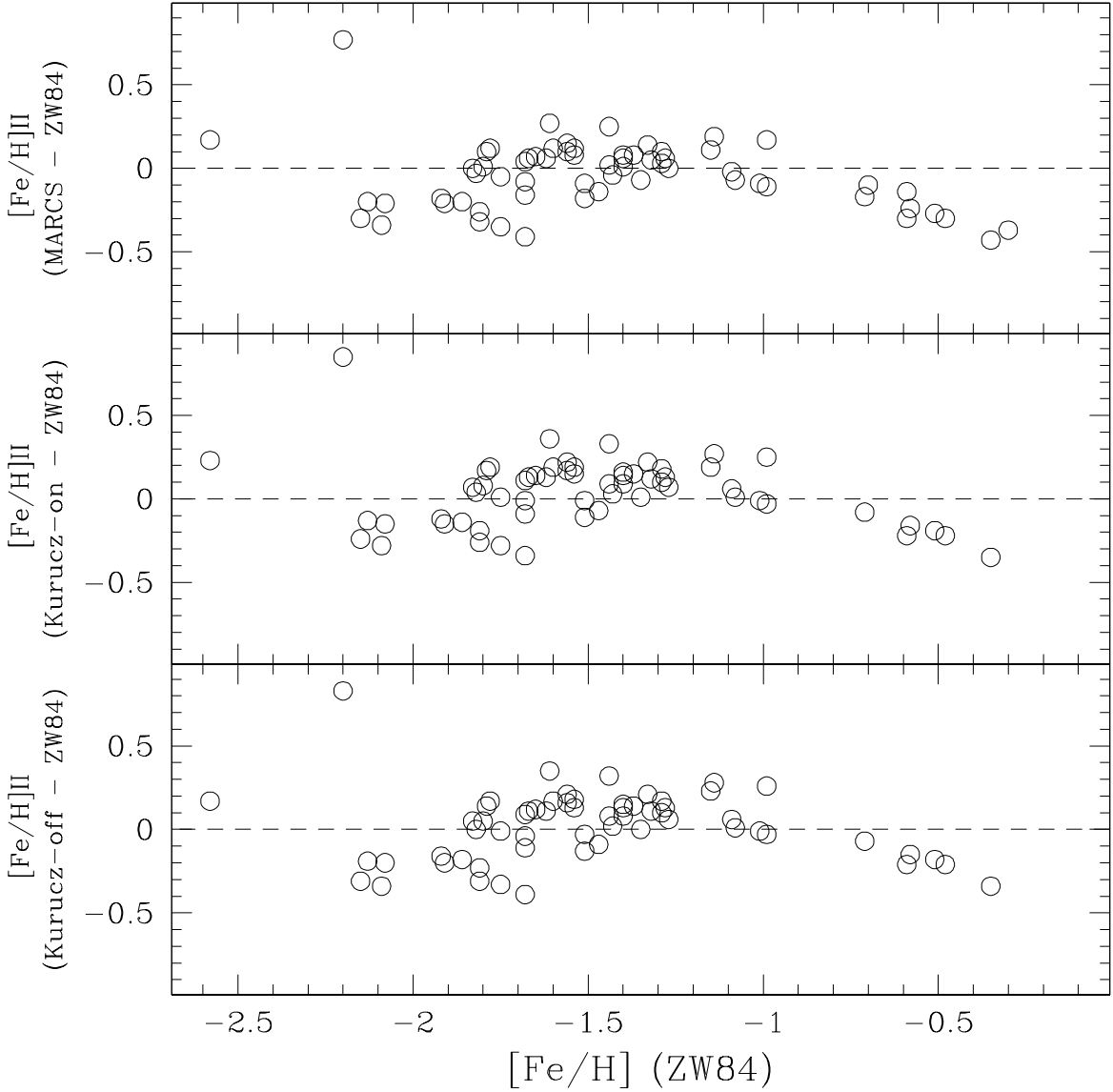}
\caption{Comparison of the metallicities derived in this study
against those derived by \citet{zw84}.  From top to bottom, the
three panels correspond to metallicities derived using MARCS 
models, Kurucz models with overshooting, and Kurucz models 
without overshooting.  Displayed are the differences as a 
function of those derived by \citet{zw84}.\label{figure3}}
\end{figure}

      By way of comparison, we plot in Figure~\ref{figure3} the 
values of [Fe/H]$_{\rm II}$ derived here for MARCS models and 
Kurucz models (with convective overshooting) and the values of 
[Fe/H] given by ZW84, taken from the tabulation of Rutledge 
{\it et al.}\ (1997b\nocite{rutledge+97b}, their Table 2, 
column 3).  Compared with Figure~\ref{figure2}, this plot has the 
additional interest that it involves many more clusters than are 
available in the more restricted CG97 sample. Whether we employ 
MARCS or Kurucz models, the qualitative shapes of the plots are 
the same: at intermediate metallicities, metallicities from ZW84 
are the same as those of MARCS Fe~{\sc ii} models and only 0.1 dex 
higher than the Kurucz Fe~{\sc ii} models, whereas they are 
significantly higher (by 0.2-0.3 dex) among both the most 
metal-poor and most metal-rich clusters of the sample.

\section{Adoption of the Houdashelt {\it et al.}\ T$_{\rm eff}$-Scale\label{6adopthoud}}

     \citet{ivans+01}, in their study of 36 giants in M5, compared 
the values of T$_{\rm eff}$ obtained from the \citet{houdashelt+00}
and \citet{alonso+99} color calibrations and found an average 
offset of +43K $\pm$ 20K, in the sense {\it Houdashelt 
{\it et al.}\ minus Alonso  {\it et al.}} The offset appeared to 
diminish with decreasing luminosity for stars in this cluster. In
the present case, we examined giants in several key clusters over 
a range of metallicities, and found that the mean offsets, in the 
same sense, ranged from 30K at [Fe/H] $\simeq$ --0.7 to as much as 
130K at [Fe/H] $\simeq$ -2.4.  These results were obtained by 
matching observed and predicted $($$V$--$K$$)_{0}$ colors.  

In Table~\ref{table5} we tabulate values of the color calibration
offsets rounded to the nearest 5K.  These apply only to the 
brighter stars in each cluster, since $V$--$K$ measurements are not 
generally available for the stars with T$_{\rm eff}$ $\gtrsim$ 4500K.

Included in Table~\ref{table5}, columns 4 and 5, are the effect 
of these offsets on the derived Fe~{\sc i} and Fe~{\sc ii} 
abundances; the color calibration offsets are, of course, 
independent of the atmospheric models chosen.  The shifts in 
abundance induced by the temperature increases are slightly 
offset by the corresponding increases in surface gravity.  
Still, the effect in the case of the lowest metallicities is 
quite substantial for Fe~{\sc i}.

If [Fe/H] based on Fe~{\sc ii} is indeed the {\em correct}
representation of the metallicity, then all metallicities 
would be driven lower by typically $\sim$0.04 dex independent 
of metallicity, should the Houdashelt {\it et al.}\ 
T$_{\rm eff}$ scale be adopted.

\section{The Question of Iron Overionization\label{7overionize}}

     TI99\nocite{ti99} predicted that in low metallicity stars
iron should suffer overionization as a result of departures from 
LTE, and that the effect should become more serious with 
decreasing metallicity (see discussion in \S~\ref{1intro}
of this study).  On the basis of the MARCS models adopted here, 
we examine this issue. The results are shown in 
Table~\ref{table6}, where the observed and predicted offsets 
between [Fe/H]$_{\rm II}$ and [Fe/H]$_{\rm I}$ are given in columns 
3 and 4. It is obvious that the predicted offsets are 
substantially more negative than we actually obtain from the 
observations plus MARCS models. If we had taken our values of 
T$_{\rm eff}$ from the Houdashelt {\it et al.}\ scale, the values 
of $\Delta$[Fe/H](obs) would have been even higher, by amounts 
ranging from +0.06 to +0.15 dex with decreasing metallicity, so
that the ``observed'' overionization of iron actually would 
become an underionization!

     We note that this result does not necessarily mean that the 
conclusions of TI99 are incorrect. It is generally conceded that 
one-dimensional (``1D'') LTE models give an accurate 
representation of the abundance of Fe~{\sc ii}, but this is not 
the case for Fe~{\sc i}. If that is the case, then there is no 
reason to {\em accept} as correct the Fe~{\sc i} abundances 
derived here, and this would be true whether we had taken MARCS 
or Kurucz models --- all are 1D models based on LTE. A set of 
``correct'' Fe~{\sc i} abundances must surely await the 
introduction of 3D models of the kind initiated in the studies 
by \citet{agp01}.

\section{The Rutledge {\it et al.}\ $<$$W'$$>$ versus [Fe/H]$_{\rm II}$ Calibration\label{8rutledge}}

     Having established metallicity scales for five key 
clusters based on Fe~{\sc ii}, we examine the extension of the 
scale to other clusters using the reduced EW of the IR 
Ca~{\sc ii} triplet, a quantity which when averaged over the 
group of giants observed in any one cluster, is denoted by 
$<$$W'$$>$ \citep{rutledge+97a, rutledge+97b}.  The values of 
$<$$W'$$>$ are listed in Table~\ref{table4}.  $<$$W'$$>$ was not 
measured for M3 and NGC 7006.  Instead, we resorted to estimating 
$<$$W'$$>$ for these clusters from a plot of the integrated 
metallicity index $Q39$ \citep{zinn80} versus $<$$W'$$>$. Thus 
for these two clusters the estimated error in $<$$W'$$>$ is 
somewhat larger than the typical error of 0.05 to 0.10 quoted for 
the key clusters by Rutledge {\it et al}.

     We plot $<$$W'$$>$ versus [Fe/H]$_{\rm II}$ for MARCS and the two 
choices of Kurucz models in Figure~\ref{figure4}, assuming the 
correctness of the Alonso {\it et al.}\ T$_{\rm eff}$ scale.  We 
do not make a separate plot based on the Houdashelt 
{\it et al.}\ T$_{\rm eff}$ scale, since as noted in 
\S~\ref{6adopthoud}, the reduction in [Fe/H]$_{\rm II}$ amounts to 
only $\sim$0.04 dex, and is independent of metallicity.

     The regressions of [Fe/H]$_{\rm II}$ on $<$$W'$$>$ are remarkably 
linear for all model choices.  The coefficients and standard errors
($\sigma$) are as follows:

\begin{enumerate}
\item 
\indent
[Fe/H]$_{\rm II}$ = 0.531$_{(\pm 0.025)}$ $\times$ $<$$W'$$>$ -- 3.279$_{(\pm 0.086)}$   (MARCS)
\item 
\indent
[Fe/H]$_{\rm II}$ = 0.537$_{(\pm 0.024)}$ $\times$ $<$$W'$$>$ -- 3.225$_{(\pm 0.082)}$  (Kurucz with conv.\ overshooting)
\item 
\indent
[Fe/H]$_{\rm II}$ = 0.562$_{(\pm 0.023)}$ $\times$ $<$$W'$$>$ -- 3.329$_{(\pm 0.078)}$ (Kurucz~without~conv.~overshooting)
\end{enumerate}

We caution the reader that there is no {\it a priori} reason 
to expect linearity in these relationships. But the linearity 
does provide a convenient method for estimating [Fe/H]$_{\rm II}$ 
from $<$$W'$$>$. On the basis of these regressions, we list in 
Table~\ref{table7} the values of [Fe/H]$_{\rm II}$ for all 
globular clusters listed by Rutledge {\it et al.}\ 
(1997b\nocite{rutledge+97b}, their Table 2), for which 
[Fe/H]$_{\rm II}$ $\lesssim$ --0.65.  

There exists a large discrepancy in the [Fe/H] value of Pal 4 
presented in Table~\ref{table7}:  ZW84 and the other estimates 
differ by $\sim$1 dex.  ZW84 based their estimate on an upward 
revision of the earlier result of --2.4 by \citet{cs78}.  Since 
then, \cite{rh86} derived a value of [Fe/H] $\sim$ --1 (on the 
basis of isochrone fits); \citet{adcz92} derived a metallicity 
on the ZW84 scale of --1.28 $\pm$ 0.20 (employing the Ca~triplet 
method); and \citet{rutledge+97b} derived --1.50 $\pm$ 0.18 on 
the ZW84 scale.  These later values reasonably bracket our 
metallicity estimates for this cluster.

\section{Metal-Rich Clusters ([Fe/H] $>$ --0.65)\label{9metal-rich}}

     So far we have dealt only with clusters having 
metallicities equal to, or more-metal poor, than M71 and 47 
Tuc. The more metal-rich clusters, owing to their concentration 
in or near the Galactic nuclear bulge, are often obscured by 
interstellar dust and thus difficult to study at the required 
high spectral resolution. The most favorable cases, NGC 6553 
and NGC 6528, have recently benefitted from high resolution 
analysis owing to the operation of the HIRES spectrograph of 
the Keck~I telescope, now to be discussed.

     Since these clusters are metal-rich, we expect that (1)
the Fe~{\sc i} excitation plot should provide an accurate 
estimate of T$_{\rm eff}$ and (2) the ionization equilibrium of 
Fe~{\sc i} and Fe~{\sc ii} should be satisfied in LTE.  It then 
follows that [Fe/H] can be estimated from [Fe/H]$_{\rm I}$ 
alone.  These are fortunate circumstances since reddening for 
these clusters is quite large ($E(B$--$V)$ $\gtrsim$ 0.50), 
uncertain, and possibly variable, thus introducing uncertainty 
in assigning T$_{\rm eff}$ and log~$g$ values solely from colors 
and absolute magnitudes. These conditions were met in recent 
studies based on spectra of resolution R $\sim$ 34000 obtained 
with the Keck~I HIRES spectrograph. For the bulge cluster NGC 
6553, \citet{cohen+99} found [Fe/H] = --0.16, later revised to 
[Fe/H] = --0.06 $\pm$ 0.15, based on analysis of five red 
horizontal branch (``HB'') stars \citep{carretta+01}. The 
difference between [Fe/H]$_{\rm I}$ and [Fe/H]$_{\rm II}$ was 
found to be negligible.  Analysis of a similar set of spectra of 
four red HB stars in NGC 6528, a well-known cluster in Baade's 
window, led to [Fe/H] = +0.07 $\pm$ 0.01 \citep{carretta+01}.

\begin{figure}
\plotone{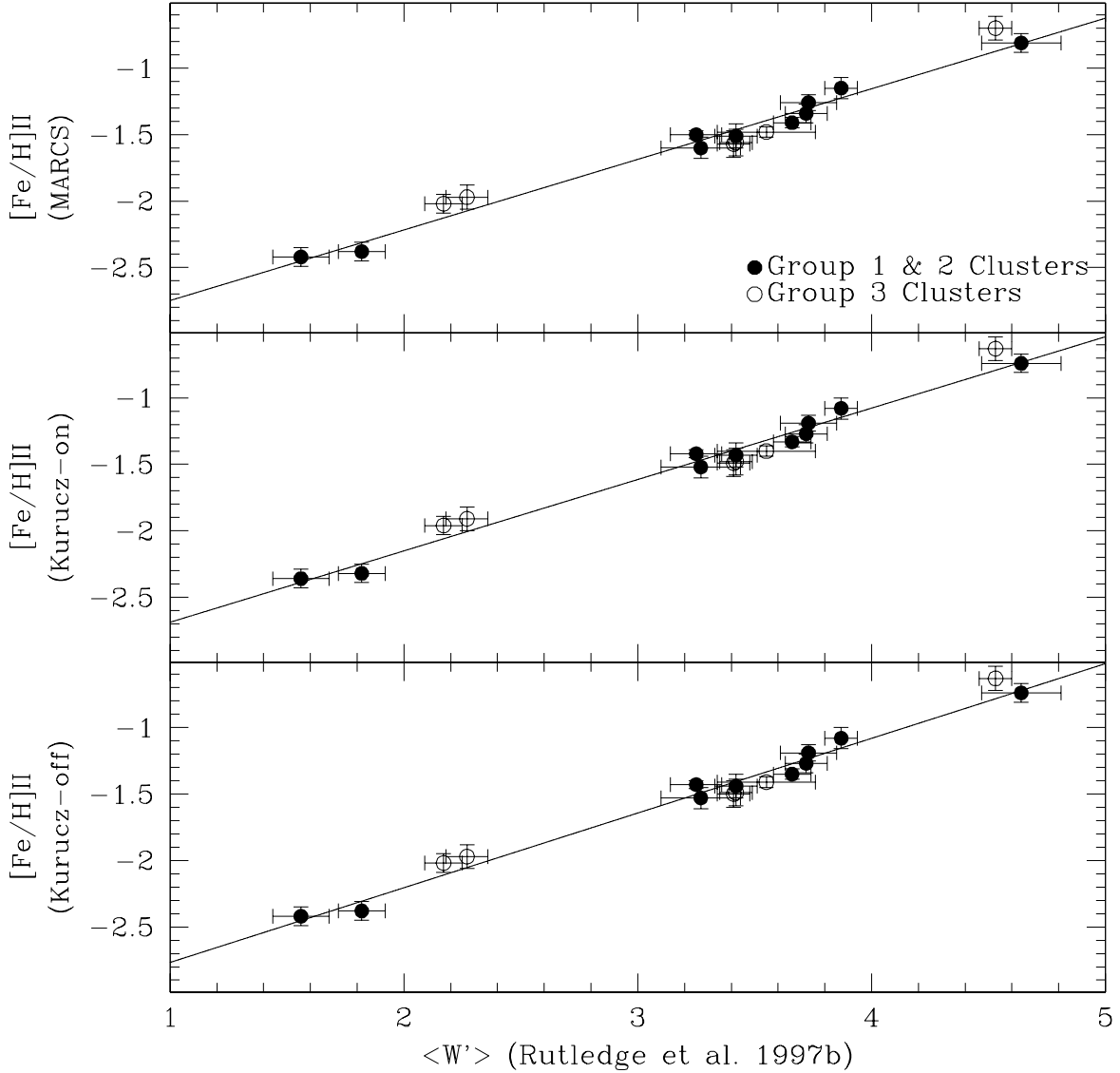}
\caption{The top panel shows our [Fe/H]$_{\rm II}$ results 
from Table~\ref{table4} employing MARCS as a function of 
$<$$W'$$>$ from \citet{rutledge+97b}; the middle panel 
displays the same information for Kurucz models with 
overshooting; the bottom panel displays the same information 
for Kurucz models without overshooting.  Symbols are those 
described in Figure~\ref{figure2}.  The solid line in each 
panel corresponds to a linear regression on $<$$W'$$>$.  The 
$<$$W'$$>$ error bars for clusters M3, M92, and NGC7006 are 
estimated from the correlations employed to derive $<$$W'$$>$ 
for those clusters (see Table~\ref{table4}).\label{figure4}}
\end{figure}

    Values of $<$$W'$$>$ for NGC 6553 and NGC 6528 are 
respectively 4.96 and 5.41 \citep{rutledge+97b}. All of the 
correlations illustrated in Figure~\ref{figure4}, if 
extrapolated, would predict lower values of [Fe/H], by 
amounts ranging from 0.3 to 0.5 dex, than actually measured. 
It therefore seems clear that the correlations are non-linear
for clusters with metallicities exceeding [Fe/H] $\simeq$ 
--0.7 and must curve ``upward''.  In part, this probably 
reflects the increasing saturation of the IR Ca~{\sc ii} 
triplet lines with increasing metallicity.  We note, however, 
that a study of two red giants in NGC 6553 by \citet{barbuy+99} 
yielded a considerably lower metallicity, [Fe/H] = --0.55, than 
was found by Cohen {\it et al.}, one that would substantially 
agree with values deduced from extrapolation of the 
correlations of Figure~\ref{figure4}. 

Nevertheless, the higher abundances are to be preferred for 
several reasons. The Cohen {\it et al.}\ results are based 
on more than twice the number of stars as Barbuy {\it et 
al.}, the stars themselves have higher T$_{\rm eff}$'s, less 
complex spectra, and therefore better defined continuum 
levels, and finally the spectra themselves have 1.5 times the 
spectral resolution. The difference in metallicity cannot be 
explained away by differences in the choice of models or 
log~{\it gf}-values. Further work on bulge clusters will no 
doubt emerge from the new large ESO, Gemini, and Magellan 
telescopes in Chile, and resolution of these differences will 
be forthcoming.

\section{The Determination of [X/Fe]-Ratios: A Brief Discussion\label{10ratios}}

     The main contention of the foregoing discussion is 
simply that if Fe is a surrogate for ``metallicity'', then 
metallicity should be represented by Fe~{\sc ii} since it 
is by far the dominant species of Fe. Use of Fe~{\sc i} for 
this purpose is also clouded by possible, if uncertain, 
departures from LTE.

It follows then that [X/Fe]-ratios should ordinarily be 
normalized to Fe abundance based on the analysis of 
Fe~{\sc ii} lines. The actual situation is, however, not 
quite so simple. Although an extensive treatment of the 
problem is beyond the scope of this paper, we briefly 
discuss the main issues, with attention focussed on the 
particular atomic or molecular features with which any 
given element manifests itself in the optical and near 
IR spectra of globular cluster giants.

\subsection{C, N, O Group\label{10ratios.cno}}

     These elements share the property that their first 
ionization potentials are very high (11.2 -- 13.6 eV) in 
comparison with the characteristic atmospheric temperatures 
of cluster giants (kT $\sim$ 0.3 to 0.6 eV), and thus are 
overwhelmingly in the neutral state. It follows that 
estimates of the amount of neutral C, N, and O should be 
normalized to the amount of Fe based on Fe~{\sc ii}. 
However, estimates of the amount of neutral C, N, or O
actually present depend crucially on which feature 
involving the element is spectroscopically accessible. The 
neutral state term-schemes of these elements are similar in 
that the available lines of the C~{\sc i}, N~{\sc i}, and 
O~{\sc i} all arise from levels of high excitation 
potential, are therefore generally weak and often suffer 
from non-LTE effects ({\it e.g.}, Mishenina {\it et al.}\ 
2000\nocite{mishenina+00}, Kiselman 
2001\nocite{kiselman01}).

The high excitation potentials also demand exquisitely 
precise knowledge of the atmospheric temperature structure 
if the line strengths are to be converted to accurate C, N, 
O abundances (Tomkin {\it et al.}\ 1992\nocite{tomkin+92}, 
King 1993\nocite{king93}, 2000\nocite{king00}). In the 
case of oxygen, this difficulty can be avoided by measuring 
the EWs of the [O~{\sc i}] lines at $\lambda\lambda$6300, 
6364\AA, which in globular cluster giants generally range 
from $\sim$5 to $\sim$40 m\AA.  In the Sun, these lines
are weak and have EWs $<$ 6 m\AA. The [O~{\sc i}] lines 
arise from the ground state, have well-known log 
{\it gf}-values, and are formed in LTE. However, recent 
analysis \citep{nissen+02}, using 3D models, indicates that 
[O/Fe]-ratios based on abundances of O from [O~{\sc i}] and 
Fe from Fe~{\sc ii} suffer a reduction of $\sim$0.2 dex 
compared with [O/Fe]-ratios based on [O~{\sc i}], Fe~{\sc ii} 
and 1D models. The [O~{\sc i}] lines almost certainly provide 
the most reliable means of estimating log $\epsilon$(O), but 
the effect of 3D modelling must be taken into account.

     Abundances of C, N, and O are also derivable from 
diatomic molecules such as CH, NH, CN, CO, and OH, all of 
which appear in the optical or near-IR region of the 
spectra of globular cluster giants. It would take us too
far afield to discuss the extensive literature making use 
of these molecules in the study of the spectra of globular 
cluster giants, but a few remarks on oxygen abundances 
derived from OH compared with [O~{\sc i}] are in order, 
since the abundance of O in old metal-poor stars is 
relevant to models of Type~II supernova ejecta, to the 
problem of determining the ages of the oldest stars, and 
to stellar evolution ({\it e.g.}, Pagel \& Tautvaisiene 
1995\nocite{pt95}, Woosley \& Weaver 1995\nocite{ww95}, 
Israelian {\it et al.}\ 1998\nocite{israelian+98}, 
Kraft 2000\nocite{kraft00}, Melendez {\it et al.}\ 
2001\nocite{melendez+01}, Sneden \& Primas 
2001\nocite{sp01}). 

     Observation and LTE analysis of the near UV OH 
electronic transition bands at 
$\lambda\lambda$3080--3300\AA\ \citep{israelian+98,
boesgaard+99} have led to values of [O/Fe] some 0.2 to 
0.5 dex higher than values derived from [O~{\sc i}] in 
field halo stars having metallicities comparable to those 
of globular clusters in the range --1.2 $\leq$ [Fe/H] 
$\leq$ --2.4. However, the amount of O tied up in OH 
renders OH a distinctly minor component of species 
involving oxygen, and the derivation of O from OH is 
quite sensitive to local variations in temperature and 
pressure in stellar atmospheres. The matter has been 
studied using 3D models by \citet{asplund01} and 
\citet{agp01}.  These authors concluded that OH formation 
is dominated by comparatively low-temperature atmospheric 
regions not considered in conventional 1D models.  The 
result is that the values of O derived from OH in 3D 
models are as much as 0.5 dex lower than what is inferred 
from 1D models.  In any case, use of OH to derive log
$\epsilon$(O) has great utility since lines of [O~{\sc i}] 
virtually disappear among metal-poor stars with surface 
gravities higher than those of giants, {\it e.g.}, halo 
field subdwarfs and main sequence dwarfs in globular 
clusters, and halo stars with metallicity lower than that 
of the most metal-poor globular clusters ([Fe/H] $\lesssim$ 
--2.5). In such stars, the UV OH bands can be studied to a 
metallicity as low as [Fe/H] $\sim$ --4 
\citep{israelian+01}.

     Studies of near IR OH rotation-vibration bands (1.5 
-- 1.7 microns) have recently been inaugurated owing to 
advances in IR detector technology \citep{bc96, 
balachandran+01, melendez+01}.  In the abundance range of 
metal-poor globular clusters, field halo (mostly) giants 
yield [O/Fe]-ratios in close agreement with values derived
from the [O~{\sc i}] lines, in sharp contrast with the 
discrepancy found from the UV OH bands (compare Israelian 
{\it et al.}\ 1998\nocite{israelian+98} with Melendez 
{\it et al.}\ 2001\nocite{mendelez+01}) suggest that the 
discordance arises from different choices of log 
{\it gf}-values for the UV OH bands. Meanwhile, the effect 
of 3D calculations on the derivation of oxygen abundances 
from the IR OH bands has not so far been considered.  Yet 
to be considered as well is the effect of 3D modelling on 
carbon and nitrogen abundances that have been derived 
previously from 1D modelling of CH, NH, CN and CO bands. 
We consider it premature to discuss here the derivation of 
C and N abundances from these diatomic molecules, until
such time as the effects of 3D modelling can be discussed.

     Finally it appears that all [O/Fe]-estimates for 
globular cluster giants having [Fe/H]$_{\rm II}$ $<$ --0.7 
need to be systematically raised. It has been shown 
\citep{reetz99, allende+01} that the solar 
$\lambda$6300\AA\ [O~{\sc i}] line is afflicted with a 
heretofore unacknowledged very weak line of Ni~{\sc i} 
arising 4 eV above the ground state. This line presumably
disappears in low-metallicity giants having temperatures 
considerably lower than that of the Sun. 
In addition, 3D modelling of the solar [O~{\sc i}] line 
also reduces the solar oxygen abundance. Adding the two 
effects results in a 0.24 dex reduction in the value of log 
$\epsilon$(O) in the Sun, from 8.93 to 8.69 $\pm$ 0.05
\citep{allende+01}. Since [O/Fe]-values are by definition 
quoted in units of the solar [O/Fe]-ratio, the total effect 
corresponds to a systematic increase of $\sim$0.24 dex in 
[O/Fe] among metal-poor globular cluster giants.

\subsection{Elements Appearing in the Neutral State\label{10ratios.neutrals}}

     Included here are Na, Mg, Al, Si, Ca and Ti among 
the so-called ``$\alpha$'' and ``odd'' elements, along with 
V, Cr, Mn and Ni among the ``Fe-peak'' elements.  Their 
first ionization potentials range from $\sim$5 to $\sim$8 
eV, and they appear in giant star spectra in the neutral 
state. If the TI99 conjecture that Fe is overionized is 
correct, then these elements would also be overionized:   
abundances derived from the neutral state will all be too 
small. However, it is not obvious that the derived 
underabundance, or ``overionization factor'', appropriate to 
Fe~{\sc i} is applicable to these elements. 

The ionization potentials of these neutral elements are not 
exactly the same as Fe, so a neutral atom in the atmosphere 
``sees'' a different photon radiation field than does an Fe 
atom. Some ionizations also occur from low lying excited 
states, but the term schemes of these atoms differ, 
sometimes radically.  Obviously, calculations similar to 
those made by TI99 would need to be made for the neutral 
state of each of these species. In the meantime, we can 
only suggest that [X/Fe] be normalized to [Fe/H]$_{\rm I}$, 
assuming that the overionization effect for these elements 
is similar to that of Fe. On the basis of LTE 1D models, 
the factor is quite small (--0.03 $\pm$ 0.05 from 
Table~\ref{table6}), but again we note that 3D models may 
tell a different story.

\subsection{Sc, Ti and Heavy Elements such as Ba, La and Eu\label{10ratios.ions}}

     Lines of Sc~{\sc ii} and Ti~{\sc ii} also appear in 
the spectra of globular cluster giants, and correspond to 
the principal stage of ionization. This is also the case 
for the heavy neutron-capture nucleosynthesis species such 
as Ba, La and Eu. In these cases, clearly the derived 
abundances must be normalized to the Fe abundance based 
on Fe~{\sc ii}.

\section{Summary\label{11summary}}

     Metallicity scales for globular clusters are normally 
set on the basis of low to medium resolution spectroscopic 
(Rutledge {\it et al.}\ 
1997a\nocite{rutledge+97a},b\nocite{rutledge+97b}) or 
metallicity-sensitive photometric \citep{zw84} indices, so 
that observations can be made of a large sample of clusters 
widely distributed in the Galaxy.  However, the indices must 
be calibrated by measuring [Fe/H] from high resolution 
spectra of giants in key nearby clusters. Most values of 
[Fe/H] have generally been derived from analysis of 
Fe~{\sc i}, a species richly endowed with lines of varying 
strength and excitation. Following these precepts, \citet{zw84}
obtained metallicities for 121 Galactic globular clusters, 
measuring metallicity sensitive photometric indices of 
integrated cluster light. Recently \citet{cg97} revised the 
ZW84 scale on the basis of improved Fe~{\sc i} abundances of 
24 key clusters. Although the rank ordering of metallicity 
among clusters was not changed by the revision, moderate 
systematic differences emerged, most notably an offset of as 
much 0.25 dex in [Fe/H] among clusters of intermediate metal 
deficiency (--1.9 $<$ [Fe/H] $<$ --1.0).

     Since the establishment of the CG97 abundance scale, 
several new developments suggest a need for further revision 
that is undertaken here.  These are: (1) the introduction of 
two new scales of T$_{\rm eff}$ versus colors for low-mass red 
giant stars \citep{alonso+99, houdashelt+00}; (2) a small 
general increase in cluster distance moduli as a result of
Hipparcos-based parallaxes of halo field subdwarfs 
\citep{reid98, carretta+00}; (3) the contention that Fe is
``overionized'' in the atmospheres of metal-poor stars 
\citep{ti99}; and (4) the deficiency of 1D models in 
providing an adequate representation of the atmospheres of 
metal-poor stars \citep{agp01}.

     Assuming that, in the atmospheres of low-mass red 
giant stars, 1D LTE models accurately predict the abundance 
of Fe from Fe~{\sc ii}, we attempt to set a globular cluster 
metallicity scale based on the EWs of Fe~{\sc ii} lines
measured from high resolution spectra of giants in 16 key 
clusters lying in the abundance range --2.4 $<$ 
[Fe/H]$_{\rm II}$ $<$ --0.7.  We base the scale largely on the 
analysis of spectra of 149 giants in 11 clusters by the 
Lick-Texas (LTG) group supplemented by other high resolution 
studies of giants in five clusters. Preliminary assumptions
included: (1) adoption of T$_{\rm eff}$ values from the 
Alonso {\it et al.}\ color versus T$_{\rm eff}$ scale, with 
preference given to T$_{\rm eff}$ based on $($$V$--$K$$)_{0}$ when 
known; (2) derivation {\it ab initio} of true distance moduli 
for five key clusters as a means of setting stellar surface 
gravities; (3) adoption of MARCS models for analysis of giant 
star spectra; (4) reconsideration of log~{\it gf}-values for 
Fe~{\sc ii} based on adoption of log $\epsilon$(Fe)$_{\sun}$ = 
7.52 and a Kurucz model without convective overshooting to 
represent the solar flux spectrum \citep{peterson+02}.  
Allowances are made for changes in the abundance scale if one 
adopts: (1) Kurucz models with and without convective 
overshooting to represent giant star atmospheres instead of 
MARCS models and (2) the Houdashelt {\it et al.}\ color versus 
T$_{\rm eff}$ scale is employed in place of the Alonso 
{\it et al.}\ scale.

     It is found that, despite uncertainties in the 
ability of 1D LTE models to represent the spectrum of 
Fe~{\sc i}, the values of T$_{\rm eff}$ derived from the 
Fe~{\sc i} excitation plot and that derived from the 
Alonso {\it et al.}\ color scale are essentially the 
same, in the case of seven clusters of small or vanishing 
color excess $E(B$--$V)$. This encourages the belief that 
for clusters of large or uncertain color excess, values 
of T$_{\rm eff}$ can be obtained, as well as estimates of 
$E(B$--$V)$, based entirely on the Fe~{\sc i} excitation 
plot. We also find that the degree of Fe 
``overionization'' predicted by TI99 is much muted, if Fe 
abundances are derived from Fe~{\sc i} lines in these 
same giant star analyses. However, this result may be 
misleading since the analysis of the Fe~{\sc i} spectrum 
is based on the assumed correctness of 1D LTE models, 
which is applicable only to the study of Fe~{\sc ii}, not 
Fe~{\sc i}.

     We find that [Fe/H]$_{\rm II}$, irrespective of whether 
one employs MARCS or Kurucz models, is correlated 
essentially linearly with $<$$W'$$>$, the reduced strength of 
the near IR Ca~{\sc ii} triplet defined by Rutledge 
{\it et al.}, although the actual correlation 
coefficients depend on the atmospheric model employed. 
Use of the Houdashelt {\it et al.}\ T$_{\rm eff}$ scale 
displaces [Fe/H]$_{\rm II}$ downward by $\sim$0.04 dex 
independent of model or metallicity. Clusters more 
metal-rich than [Fe/H] $\sim$ --0.65 appear to have
metallicities higher than simple extrapolations of these 
correlations would allow ({\it c.f.}\ Carretta 
{\it et al.}\ 2001\nocite{carretta+01}). The fact that 
these correlations are linear is a convenience, and may
not indicate the existence of some fundamental 
relationship. 

The metallicities derived here, based on Fe~{\sc ii}, in 
general run about 0.2 dex lower than those of CG97. They 
agree quite well with those of ZW84 in the range --1.7 $<$ 
[Fe/H] $<$ --1.0 using MARC models, but run 0.2 -- 0.3 dex 
lower at both higher and lower metallicities.

     We have discussed how to estimate [X/Fe]-ratios. 
Since C, N, and O are surely overwhelmingly in the neutral 
state, their abundances should be be determined, if possible, 
from direct indicators of the abundance of the neutral 
species, and should be normalized to the abundance of iron 
based on Fe~{\sc ii}. We discuss briefly the case of oxygen, 
noting the controversy concerning the abundance differences 
found from analysis of [O~{\sc i}], the permitted O~{\sc i} 
near-IR triplet, the near-UV OH bands and the IR OH bands. 
We recommend that log $\epsilon$(O) be derived from 
[O~{\sc i}] where possible, normalized to [Fe/H] based on
Fe~{\sc ii}, and should take into account the recent decrease 
in the assumed solar oxygen abundance \citep{allende+01}. The 
higher oxygen abundances derived from OH are likely a result 
of inadequacy of 1D models to represent the spectrum of OH, 
and may be correctable by adoption of 3D models (Asplund \&
Garcia Perez 2001\nocite{agp01}; but see also Nissen 
{\it et al.}\ 2002\nocite{nissen+02} for further discussion). 
Elements such as Ti, Sc, Ba, La and  Eu which are represented 
in the spectra by lines in the principal state of ionization, 
in this case the singly ionized state, should have their [X/Fe] 
values directly normalized to the Fe abundance based on 
Fe~{\sc ii}.  For other elements, mostly $\alpha$- and 
odd-element and Fe-peak species, they too are largely in the 
singly ionized state in the atmospheres of metal-poor giants, 
but are represented in the spectra in the neutral state. Here 
we recommend, as a temporizing measure, normalization to [Fe/H] 
based on Fe~{\sc i}, assuming that the possible ``overionization'' 
of these elements is similar to that of Fe. Further studies 
are needed to evaluate the reliability of this assumption 
or if these species, as well as Fe, indeed suffer 
significant overionization of the kind envisaged by TI99.

     We finally make a general observation about the 
globular cluster metallicity scale. We list in 
Table~\ref{table7} values of [Fe/H]$_{\rm II}$ for 68 
globular clusters based on the values of $<$$W'$$>$ and 
the three correlations based on MARCS and Kurucz models 
with and without convective overshoot.  In addition, we 
note again that T$_{\rm eff}$'s based on Alonso {\it et al.}\ 
were employed; for the Houdashelt {\it et al.}\ scale, 
0.04 dex needs to be subtracted from [Fe/H]$_{\rm II}$ for all 
correlations. 

The conclusion we reach is that there exists no definitive 
set of cluster metallicities that are systematically 
reliable on the 0.02 to 0.05 dex level. Any discussion 
using cluster abundances needs to state clearly the 
underlying assumptions concerning models used, whether 
Fe~{\sc i} or Fe~{\sc ii} or a mean thereof is what is 
meant by ``metallicity'', which T$_{\rm eff}$ scale has been 
adopted, {\it etc.}  If systematic effects on the 0.25 dex 
level are irrelevant to the discussion, then almost 
anybody's metallicity scale is acceptable.

\acknowledgments

This research has made use of the NASA/IPAC Extragalactic Database 
which is operated by the Jet Propulsion Laboratory, California 
Institute of Technology, under contract with NASA; and has made 
use of NASA's Astrophysics Data System Bibliographic Services.  
Research by III was generously supported in part by McDonald 
Observatory and Continuing Fellowships from The University of Texas 
at Austin; and happily, is currently supported by NASA through 
Hubble Fellowship grant HST-HF-01151.01-A from the Space Telescope 
Science Institute, which is operated by the Association of 
Universities for Research in Astronomy, Incorporated, under NASA 
contract NAS5-26555.

We gratefully acknowledge helpful discussion with Raffaele Gratton,
Ruth Peterson, Jon Fulbright, Martin Asplund, Guy Worthey and Don 
VandenBerg.  Chris Sneden and Mike Bolte kindly read a first draft 
of the manuscript and provided useful comments. We thank Roger Bell 
for the results of a highly relevant calculation, and Matt Shetrone 
for conveying unpublished results.



\appendix

\section{Revised Cluster Abundances Based on Fe~{\sc ii}\label{Arevabunds}}

     We take up each of the Groups 1 to 3, and within each 
group discuss the clusters in order of decreasing 
metallicity.

\subsection{Group 1 clusters: T$_{\rm eff}$ and log~$g$ based on colors and absolute magnitudes\label{Arevabunds.g1}}

\subsubsection{M5 (NGC 5904); $E(B$--$V)$ = 0.03, (m--M)$_{0}$ = 14.42\label{Arevabunds.g1.m5}}
         
        The 36 giants analyzed by LTG made use of the 
techniques employed in this paper (see Ivans {\it et al.}\ 
2001\nocite{ivans+01}, their Tables 5, 6 \& 8) so results are 
not repeated in detail here. The only adjustments we make are 
those driven by the increases of Fe~{\sc ii} and Fe~{\sc i} 
log {\it gf}-values: --0.05 in [Fe/H]$_{\rm II}$ and --0.02 
in [Fe/H]$_{\rm I}$. We thus obtain $<$[Fe/H]$_{\rm II}>$ = 
--1.26 ($\sigma$ = 0.06) and $<$[Fe/H]$_{\rm I}>$ = --1.36 
($\sigma$ = 0.07).  The MARCS models lead to an Fe~{\sc i} 
underabundance of --0.10 dex.

\subsubsection{NGC 362; $E(B$--$V)$ = 0.05, (m--M)$_{0}$ = 14.70\label{Arevabunds.g1.n362}}

       Values of $V$--$K$ are known \citep{frogel+83} for 
almost all of the giants studied by \citet{sk00}.  We based 
our revision on the seven giants having T$_{\rm eff}$ $\geq$ 
4000K. Estimates of T$_{\rm eff}$, log~$g$, [Fe/H]$_{\rm I}$ 
and [Fe/H]$_{\rm II}$ are presented in 
Table~\ref{tableA1}.\footnote{
Abundances for Fe~{\sc i} and Fe~{\sc ii} were not 
individually listed by \citet{sk00}, but rather only the 
mean value. Separate values for both NGC 362 and NGC 288 
were kindly communicated by Dr. Shetrone for use in this 
study.}

Mean values of the abundances are $<$[Fe/H]$_{\rm II}>$ = 
--1.34 ($\sigma$ = 0.07) and $<$[Fe/H]$_{\rm I}>$ = --1.31 
($\sigma$ = 0.03).  In this case, we find a slight 
Fe~{\sc i} overabundance (see further discussion in 
\S~\ref{7overionize}). From its position in the 
Hertzsprung-Russell diagram, Star 77 is probably an AGB star, 
and we assigned to it a mass of 0.60~M$_{\sun}$, allowing for 
a typical mass loss of 0.20~M$_{\sun}$ for a star evolving 
from the RGB to the AGB.

\subsubsection{NGC 288; $E(B$--$V)$ = 0.03, (m--M)$_{0}$ = 14.66\label{Arevabunds.g1.n288}}

     Only three stars (20c, 245, 231) have known values of 
$V$--$K$ \citep{frogel+81, frogel+83}, so most T$_{\rm eff}$'s are 
derived from $($$B$--$V$$)_{0}$ based on photometry from
\citet{olszewski+84}.  Estimates of T$_{\rm eff}$, log~$g$, 
[Fe/H]$_{\rm I}$ and [Fe/H]$_{\rm II}$ are presented in 
Table~\ref{tableA2}.   We find $<$[Fe/H]$_{\rm II}>$ = 
--1.41 ($\sigma$ = 0.04) and  $<$[Fe/H]$_{\rm I}>$ = --1.36 
($\sigma$ = 0.07), once again with a small Fe~{\sc i} 
overabundance of 0.04.

\subsubsection{M3 (NGC 5272); $E(B$--$V)$ = 0.01, (m--M)$_{0}$ = 15.02\label{Arevabunds.g1.m3}}

     \citet{sneden+02} have analyzed 23 giants on the basis 
of spectra taken with the Keck~I HIRES spectrograph, using 
the techniques favored in the present study. Details will 
be found in a paper now in preparation.  Among these 23 
stars are the four giants studied in connection with the 
strong Li line found in one of them, IV-101 
\citep{kraft+99}. In Table~\ref{tableA3} we list the 
original and revised parameters for these four stars. The 
star mistakenly identified as VZ~746 by Kraft {\it et al.}\ 
is actually VZ~729.

The Fe abundances averaged over all 23 stars are 
$<$[Fe/H]$_{\rm II}>$ = --1.50 ($\sigma$ =0.03) and 
$<$[Fe/H]$_{\rm I}>$ = --1.58 ($\sigma$ = 0.06), with a 
corresponding Fe~{\sc i} underabundance of --0.08 dex.

\subsubsection{M13 (NGC 6205); $E(B$--$V)$ = 0.02, (m--M)$_{0}$ = 14.42\label{Arevabunds.g1.m13}}

     A catalog of high-resolution analyses of 34 giants found in 
\citet{kraft+97}, from which we selected 28 stars, omitting a 
few in which the S/N of the spectra was less than ideal. We 
entered the \citet{alonso+99} color versus T$_{\rm eff}$ scale using 
$($$V$--$K$$)_{0}$ where known \citep{cohen+78}, otherwise using
$($$B$--$V$$)_{0}$ \citep{cm79}.  The 28 stars range 3900K $<$ 
T$_{\rm eff}$ $<$ 5050K and --3.7 $<$ M$_{bol}$ $<$ --0.4, about 0.5 
mag above the horizontal branch. Table~\ref{tableA4} lists the 
original and revised values of T$_{\rm eff}$ and log~$g$ along with 
the revised values of [Fe/H]$_{\rm II}$ and [Fe/H]$_{\rm I}$.

  The mean Fe abundances are $<$[Fe/H]$_{\rm II}>$ = --1.60 
($\sigma$ = 0.08) and $<$[Fe/H]$_{\rm I}>$ = --1.63 
($\sigma$ = 0.06), with a corresponding small Fe~{\sc i} 
underabundance of --0.03 dex.

\subsubsection{M92 (NGC 6341); $E(B$--$V)$ = 0.02, (m--M)$_{0}$ = 14.75\label{Arevabunds.g1.m92}}

     We give preference to the analysis of three giants by 
\citet{langer+98}, which was based on many more Fe~{\sc i} 
and Fe~{\sc ii} lines than were reported by 
\citet{sneden+91}.  And, although both data sets were 
derived from Lick 3.0-m Hamilton echelle spectra, the 
Langer {\it et al.}\ spectra were obtained with the 
post-1995 improved version of the spectrograph, with 
upgraded optics and a 2048$\times$2048 pixel chip that
replaced the 800$\times$800 TI chip.  

We also include the three stars X-49, B-19, and B-95 
studied by \citet{shetrone96}, also derived from the 
post-1995 version of the Hamilton spectrograph.  The latter 
study makes use of the {\it gf}-values employed by LTG, 
and consequently all Fe~{\sc ii}-based and Fe~{\sc i}-based 
abundances were decreased by 0.05 dex and 0.02 dex, 
respectively. The investigation of \citet{langer+98} is 
based on {\it gf}-values taken from \citet{lambert+96}, 
which are essentially on the same system of 
{\it gf}-values as those of the Blackwell group, 
{\it etc.},  and thus no corrections were applied to 
the abundances. 

Colors and magnitudes for B-19 and B-95 were taken from 
the compilation by \citet{rees92}, but for the remaining 
four stars, we used the CCD photometry of \citet{bolte01}, 
some of which was reported in \citet{langer+98}.  

In Table~\ref{tableA5}, we list the original and revised 
Fe abundances.  These lead to mean values, averaged over 
the six stars, of $<$[Fe/H]$_{\rm II}>$ = --2.38 
($\sigma$ = 0.07) and $<$[Fe/H]$_{\rm I}>$ = --2.50 
($\sigma$ = 0.12) with a corresponding Fe~{\sc i} 
underabundance of --0.12 dex.  Presumably part of the 
scatter in the mean values of [Fe/H]$_{\rm II}$ and 
[Fe/H]$_{\rm I}$ is due to the variation in metallicity 
($\sim$0.1 dex) discovered by \citet{langer+98}.

     On the basis of CCD spectra of two M92 giants
(III-13 and XII-8), \citet{peterson+90} derived 
$<$[Fe/H]$>$ = --2.52.  Lines of both Fe~{\sc i} and
Fe~{\sc ii} were measured, and surface gravities 
obtained on the assumption that [Fe/H]$_{\rm I}$
= [Fe/H]$_{\rm II}$.  Their analysis shares in common 
with LTG the use of MARCS models.  On average, their
T$_{\rm eff}$ scale is very close to that of LTG, but their
values of log~$g$ are lower, by about 0.1 dex.  Again in 
common with LTG, they adopted log~{\it gf}-values of
\citet{blackwell+80}, but normalized to a solar 
log $\epsilon$(Fe) = 7.63.  Renormalizing to the solar 
log $\epsilon$(Fe) abundance adopted here would raise 
their value of [Fe/H] for M92 by 0.11 dex, bringing it into 
good agreement with the Fe~{\sc ii} result suggested here.

     \citet{king+98} also obtained [Fe/H] = --2.52 from
analysis of the EWs of Fe~{\sc i} lines in the spectra of 
six M92 subgiants.  However, no lines of Fe~{\sc ii} were
measured, so that the procedures employed here cannot be
duplicated.

\subsubsection{M15 (NGC 7078); $E(B$--$V)$ = 0.10, (m--M)$_{0}$ = 15.25\label{Arevabunds.g1.m15}}

     The list of M15 giants having measured values of $V$--$K$ 
is rather small \citep{frogel+83}, so we fall back on 
photoelectrically calibrated photographic $V$ and $B$--$V$ 
values of \citet{sandage70} and \citet{cudworth76}.  Some 
of the latter are a bit uncertain owing to the photographic 
saturation of images of the brightest stars \citep{cudworth96}. 
However, most of these were not observed by Sandage. 
Consequently we consider only stars with $V$ $>$ 13.0.  Nine 
such M15 giants are listed in Table~\ref{tableA6}, the original 
T$_{\rm eff}$ and log~$g$ values being taken from Sneden 
{\it et al.}\ (1997\nocite{sneden+97}, 2000\nocite{sneden+00}), 
all spectra having been obtained with the Keck~I HIRES 
spectrograph.

Averaging over these nine giants, we find 
$<$[Fe/H]$_{\rm II}>$ = --2.42 ($\sigma$ = 0.07)
and $<$[Fe/H]$_{\rm I}>$ = --2.50 ($\sigma$ = 0.11), so 
that the Fe~{\sc i} underabundance is --0.08 dex.

     The reddening of M15, it should be noted, is 
somewhat controversial, with values of $E(B$--$V)$ ranging 
in the literature (reviewed by Reid 1997\nocite{reid97}) 
from 0.07 to 0.13. However, a 0.03 mag change in 
$E(B$--$V)$ away from 0.10 has not a very significant 
effect on our derived values of the iron abundance. This 
corresponds to a change of T$_{\rm eff}$ by $\sim$30K, which 
changes Fe~{\sc i} by 0.04 dex but Fe~{\sc ii} by only 
0.01 dex (Sneden {\it et al.}\ 1997\nocite{sneden+97}, 
their Table 5).  We note that the value of $E(B$--$V)$
adopted here as shown in Table~\ref{table2} is virtually 
identical to that derived by \citet{schlegel+98}.

\subsection{Group II Clusters: Large and/or Uncertain Reddening\label{Arevabunds.g2}}

\subsubsection{M71 (NGC 6838); $E(B$--$V)$ = 0.32, (m--M)$_{0}$ = 12.83\label{Arevabunds.g2.m71}}

    Estimates of $E(B$--$V)$ occurring in the literature were 
discussed by \citet{frogel+79}, who found values ranging 
from 0.21 to 0.32. The authors adopted 0.25 as did 
\citet{harris96} in his June 22, 1999 compilation. If our 
contention is correct that values of T$_{\rm eff}$ derived from 
the Fe~{\sc i} excitation plot and from colors are 
essentially the same in little reddened clusters, then we 
derive $E(B$--$V)$ = 0.32 (see Table~\ref{table2}), a value
nearly identical with that derived from the 
\citet{schlegel+98} dust maps.\footnote{
The agreement may be fortuitous. The \citet{schlegel+98} 
absorption estimates refer to a value of total absorption 
outside the Galaxy, and may not be applicable to such a 
low latitude object as M71.}  
We take (m--M)$_{0}$ = 
12.83, assuming that +0.07 must be added to the modulus 
quoted in the June 22, 1999 revision of \citet{harris96} 
(see \S~\ref{3adjustments.gmass}).  The revised photometry 
of \citet{cudworth85} was adopted, replacing the earlier 
photometry of \citet{ah79}, and the $V$ magnitudes used by 
Frogel {\it et al.}\ were appropriately corrected prior 
to evaluating T$_{\rm eff}$ from $($$V$--$K$$)_{0}$.  

In the case of M71, most of the 10 giants studied by 
\citet{sneden+94} from high resolution spectra had 
measured values of $V$--$K$ from Frogel {\it et al.}  Sneden 
{\it et al.}\ took $E(B$--$V)$ = 0.30 and (m--M)$_{0}$ = 13.0. 
The present Fe~{\sc i} and Fe~{\sc ii} abundances for each 
star are based on $E(B$--$V)$ = 0.32 and (m--M)$_{0}$ = 12.83. 
The original \citep{sneden+94} iron abundances based on the 
Fe~{\sc i} excitation plot, $E(B$--$V)$ = 0.30 and (m--M)$_{0}$
= 13.0, together with these newly derived Fe abundances, 
are given in Table~\ref{tableA7}, corrected for the offsets 
in log~{\it gf}-values described earlier.

The mean Fe abundances are: $<$[Fe/H]$_{\rm II}>$ = --0.81 
($\sigma$ = 0.07) and $<$[Fe/H]$_{\rm I}>$ = --0.82 
($\sigma$ = 0.05), with essentially no offset between 
Fe~{\sc i} and Fe~{\sc ii}. The corresponding values taking 
the original input parameters from Sneden {\it et al.}\ 
are $<$[Fe/H]$_{\rm II}>$ = --0.84 ($\sigma$ = 0.06) and 
$<$[Fe/H]$_{\rm I}>$ = --0.81 ($\sigma$ = 0.05).  The small 
changes mostly reflect the $\sim$0.2 dex change in assumed 
distance modulus, as expected.

     Recently, a more extensive study involving 25 M71 
stars observed with the HIRES spectrograph at the Keck~I 
telescope has been published \citep{cohen+01, ramirez+01,
rc02}.  The stars studied cover a much larger spread in 
luminosity than is found in the LTG sample, ranging from 
the red giant tip to the main sequence turnoff. Although 
the choice of log {\it gf}-values for Fe~{\sc i} and 
Fe~{\sc ii} lines\footnote{
The number of Fe~{\sc ii} lines measured per giant in all 
cases but two lay between 5 and 9.} 
is essentially the same 
in both the Sneden {\it et al.}\ and Cohen/Ram\'irez 
{\it et al.}\ investigations, they differ somewhat in 
choices of T$_{\rm eff}$ scales and cluster reddening, as well 
as model atmospheres. Ram\'irez {\it et al.}\ adopted 
$E(B$--$V)$ = 0.25 and (m--M)$_{0}$ = 12.96, along with 
estimates of T$_{\rm eff}$ based on the T$_{\rm eff}$ versus 
$($$V$--$K$$)_{0}$ and $($$V$--$I$$)_{0}$ scales of Houdashelt 
{\it et al.} They found excellent agreement on the average 
between T$_{\rm eff}$ derived from the photometric indices and 
T$_{\rm eff}$ derived from the Fe~{\sc i} excitation plot. 

     Over the T$_{\rm eff}$ range common to the giants in LTG
and Cohen/Ram\'irez {\it et al.} studies,\footnote{
Over the common range of T$_{\rm eff}$ and log~$g$ in these
two investigations, the number of giants analyzed is
comparable:  10 in LTG and 8 in Cohen/Ram\'irez {\it et al.}}
the Houdashelt {\it et al.}\ and Alonso {\it et al.}\ 
T$_{\rm eff}$'s are in reasonable agreement for a common 
choice of $E(B$--$V)$. Thus for $E(B$--$V)$ = 0.25, the 
Ram\'irez T$_{\rm eff}$'s average only 20K hotter the LTG 
T$_{\rm eff}$'s based on the Alonso {\it et al.}\ $($$V$--$K$$)_{0}$ 
scale. (The differences would be greater for lower luminosity 
stars, but these do not figure in the LTG study). Despite the 
differences, the final abundances derived in these two 
studies are in very good agreement. Ram\'irez {\it et al.}\ 
find $<$[Fe/H]$_{\rm II}>$ = --0.84 $\pm$ 0.12 (compared 
with the revised LTG value, above, of --0.81 $\pm$ 0.07) 
and $<$[Fe/H]$_{\rm I}>$ = --0.71 $\pm$ 0.08 (compared 
with the revised LTG value of --0.82 $\pm$ 0.05). The 
derived Fe~{\sc ii} abundances differ by a negligible 
amount.

\subsubsection{M4 (NGC 6171); $E(B$--$V)$ = 0.33$_{variable}$, (m--M)$_{0}$ = 11.61\label{Arevabunds.g2.m4}}

     M4 presents a particularly difficult case if one is 
to use colors and absolute magnitudes to estimate 
T$_{\rm eff}$ and log~$g$: reddening is large and highly 
variable across the face of the cluster, and the ratio of 
total to selective absorption is anomalous ({\it e.g.}, 
Dixon \& Longmore 1993\nocite{dl93}, Lyons {\it et al.}\ 
1995\nocite{lyons+95}, Ivans {\it et al.}\ 
1999\nocite{ivans+99}).  Ivans {\it et al.}\ found good 
agreement between the values of T$_{\rm eff}$ and log~$g$ 
derived from colors and absolute magnitudes and from the 
Fe~{\sc i} excitation plot and true distance modulus of 
11.61, which assumes a cluster distance of 2.1~kpc and 
A$_{V}$ = 1.48. Considerable uncertainty exists in 
estimates of both distance and A$_{V}$ (reviewed by Dixon 
\& Longmore and Ivans {\it et al.}), but most authors 
agree that $<E(B$--$V)>$ is near 0.33, with considerable 
scatter ranging from 0.25 to 0.42. 

We re-derived values of T$_{\rm eff}$ for each star based on 
the $($$B$--$V$$)_{0}$ estimates of Ivans {\it et al.}\ plus 
the Alonso {\it et al.}\ $($$B$--$V$$)_{0}$ versus 
T$_{\rm eff}$ calibration, and we then compared the values of 
log~$g$$_{spectroscopic}$, derived from the traditional 
Fe~{\sc i} versus Fe~{\sc ii} LTE ionization equilibrium, 
with log~$g$$_{evolutionary}$ based on these new T$_{\rm eff}$
values and the true distance modulus given above.  We 
found $<$log~$g$(evol)$>$ -- $<$log~$g$(spec)$>$ = 0.00 
($\sigma$ = 0.12) averaged over 24 giants ranging 3800K
$<$ T$_{\rm eff}$ $<$ 4775K (see Ivans {\it et al.}\ 
1999\nocite{ivans+99}, their Table 2). The corresponding 
difference in T$_{\rm eff}$ in the same sense is +21K 
($\sigma$ = 58K). Ivans {\it et al.}\ found 
$<$[Fe/H]$_{\rm II}>$ = $<$[Fe/H]$_{\rm I}>$ = --1.18,
by construction, and our revision leads to 
$<$[Fe/H]$_{\rm II}>$ = --1.19 ($\sigma$ = 0.08) and 
$<$[Fe/H]$_{\rm I}>$ = --1.16 ($\sigma$ = 0.05), the 
small offset being consistent with the small change in 
T$_{\rm eff}$.

    If for the same color excess and value of A$_{V}$ we 
had adopted the geometrical distance (defined as the 
spread in proper motions versus the spread in radial 
velocities; Peterson {\it et al.}\ 
1995\nocite{peterson+95}), the values of log~$g$(evol) 
would have increased by 0.17. This in turn would lead to 
$<$[Fe/H]$_{\rm II}>$ = --1.11 and $<$[Fe/H]$_{\rm I}>$ 
= --1.14. The changes are in fact quite modest. If we 
take smaller values of A$_{V}$, the values of log~$g$ 
become smaller, which takes us back toward the previous 
values of the iron abundance. Since the effects 
consistent with non-LTE cited in this paper no longer 
constrain equality between log~$g$(spec) (based on the 
equilibrium of Fe~{\sc i} and Fe~{\sc ii}) and 
log~$g$(evol), we cannot make an {\it a priori} choice 
between these estimates. For this reason, we take a mean
and assign $<$[Fe/H]$_{\rm II}>$ = --1.15 and 
$<$[Fe/H]$_{\rm I}>$ = --1.15.

\subsubsection{M10 (NGC 6254); $E(B$--$V)$ = 0.24, (m--M)$_{0}$ = 13.41\label{Arevabunds.g2.m10}}

     The June 22, 1999 version of Harris's (1996)\nocite{harris96} 
compilation gives $E(B$--$V)$ = 0.28 and therefore (m--M)$_{0}$ = 
13.21, but as we noted in \S~\ref{3adjustments.teffred}, agreement 
between T$_{\rm eff}$'s based on $($$B$--$V$$)_{0}$ or 
$($$V$--$K$$)_{0}$ and T$_{\rm eff}$'s based on the Fe~{\sc i} 
excitation plot suggests a reddening of $E(B$--$V)$ = 0.24, and 
this in turn yields (m--M)$_{0}$ = 13.34, which becomes 13.41 with 
our empirical correction to the Harris moduli.  We have 
already noted the somewhat disquieting discrepancy between 
our value of $E(B$--$V)$ and the value derived from the 
\citet{schlegel+98} dust maps, but retain our smaller 
value to maintain internal self-consistency. 

In  Table~\ref{tableA8} we list the values of T$_{\rm eff}$ 
and log~$g$ based on the Fe~{\sc i} excitation plot and
(m--M)$_{0}$ = 13.8 \citep{kraft+95} for nine giants in 
M10,\footnote{
We omit from consideration the five stars B through G also 
studied by \citet{kraft+95} for which $B$ and $I$ are known, 
but not $V$. We judged that estimates of T$_{\rm eff}$ based on the 
rather uncertain estimates of $V$ made in that study were not 
sufficiently reliable for present purposes.} 
along with the 
present values of T$_{\rm eff}$ and log~$g$ based on colors and 
absolute magnitudes consistent with $E(B$--$V)$ = 0.24 and 
(m--M)$_{0}$ = 13.41.  Corresponding estimates of 
[Fe/H]$_{\rm II}$ and [Fe/H]$_{\rm I}$ are also given.

The newly derived mean abundances are 
$<$[Fe/H]$_{\rm II}>$ = --1.51 ($\sigma$ = 0.10) and 
$<$[Fe/H]$_{\rm I}$$>$= --1.50 ($\sigma$ = 0.05), as 
contrasted with the ``original'' values of --1.53 for 
both Fe~{\sc i} and Fe~{\sc ii}. The temperature 
offset between present values and that derived from 
the Fe~{\sc i} excitation plot is only 32K ($\sigma$ 
= 51K), as expected; the adjustments to the Fe 
abundances conform essentially to the modification 
of the distance modulus.

\subsection{Other Clusters; Other Sources\label{Arevabunds.g3}}

\subsubsection{Some Transformation Problems\label{Arevabunds.g3.probs}}

     In this section, we deal with metallicity 
determinations for clusters not studied by LTG in addition to
LTG's study of giants in NGC~7006. We consider only clusters 
in which at least three giants were analyzed and in which 
metallicities could be determined from lines of Fe~{\sc ii}. 
We also restrict attention to studies in which EWs were 
measured from modern CCD detectors. Employment of CCDs 
generally leads to good agreement in EW measurements between 
different investigators (see {\it e.g.}, Sneden 
{\it et al.}\ 1991\nocite{sneden+91}, their Figure 3 and 
Minniti {\it et al.}\ 1993\nocite{minniti+93}, their Figure 
2). 

We make adjustments in [Fe/H]$_{\rm II}$ and [Fe/H]$_{\rm I}$ 
to account for small differences in log {\it gf}-values 
chosen by a particular investigation and those chosen by LTG. 
More serious are abundance changes induced by the choice of 
atmospheric models. Whereas LTG chose MARCS models, most of 
the other investigators employed Kurucz models with 
convective overshooting. 

As discussed in \S~\ref{5transform}, we investigated the 
situation regarding model choices by analyzing some representative 
giants using different models. Here, we applied abundance offsets 
to bring all stars onto the system defined by the MARCS models.

\subsubsection{47 Tuc (NGC 104); $E(B$--$V)$ = 0.04, (m--M)$_{0}$ = 13.32\label{Arevabunds.g3.47tuc}}

     Relatively few giants have been studied in this 
well-known mildly metal-poor, little-reddened cluster, for 
which the value of $E(B$--$V)$ = 0.04, independently cataloged 
in the June 22, 1999 revision of \citet{harris96}, agrees 
perfectly with the value based on the \citet{schlegel+98} 
dust maps.  CCD-based EWs are available for four giants 
studied by \citet{bw92}, one by \citet{ndc95} and three by 
CG97\nocite{cg97}.\footnote{
We do not include here the two 47 Tuc giants transformed 
by CG97\nocite{cg97} from lower resolution non-CCD 
observations. No EWs for Fe~{\sc ii} lines were reported 
for these stars. Their inclusion has little effect on 
the mean value of [Fe/H]$_{\rm I}$ derived here.}

We re-derived the T$_{\rm eff}$ and log~$g$ values on the 
basis of $($$V$--$K$$)_{0}$ \citep{frogel+81}, adopting $E(B$--$V)$ = 
0.04 and (m--M)$_{0}$ = 13.32.  Compared with 
\citet{bw92}, our T$_{\rm eff}$'s are 0 to 25K lower and our 
log~$g$'s are typically 0.1 dex lower. In comparison with 
CG97\nocite{cg97}, our T$_{\rm eff}$'s are the same and our 
log~$g$'s are again 0.1 lower. We adjusted all log 
{\it gf}-values for Fe~{\sc i} and Fe~{\sc ii} to 
conform to those of Blackwell {\it et al.}, as explained 
earlier. \citet{bw92} employed BGEN models (essentially 
identical to MARCS models), so that no ``model adjustment'' 
was needed to place their analysis on the same basis as that 
of LTG. 

We obtained $<$[Fe/H]$_{\rm II}>$ = --0.70 ($\sigma$ = 
0.05) and $<$[Fe/H]$_{\rm I}>$ = --0.69 ($\sigma$ = 0.04) 
for the four Brown \& Wallerstein\nocite{bw92} giants, 
based on five to six Fe~{\sc ii} lines (and numerous 
Fe~{\sc i} lines). A similar treatment of the three stars 
studied by CG97\nocite{cg97}, but using Kurucz models, 
yielded $<$[Fe/H]$_{\rm II}>$ = --0.70 ($\sigma$ = 0.03) 
and $<$[Fe/H]$_{\rm I}>$ = --0.61 ($\sigma$ = 0.04).  A 
similar treatment of the one star studied by 
\citet{ndc95}, a star also studied by CG97\nocite{cg97},
yielded [Fe/H]$_{\rm II}$ = --0.56 and [Fe/H]$_{\rm I}$ = 
--0.63. The agreement is satisfactory considering that 
more than 40 Fe~{\sc i} lines were measured by 
CG97\nocite{cg97} and more than 60 by \citet{ndc95}, but 
at most only 3 Fe~{\sc ii} lines were measured in each of 
these investigations.

     Adjustments to the MARCS models (see 
\S~\ref{5transform}) then leads to $<$[Fe/H]$_{\rm II}>$ = 
--0.69 ($\sigma$ = 0.07) and $<$[Fe/H]$_{\rm I}>$ = 
--0.61 ($\sigma$ = 0.01) as the weighted mean of these 
last two investigations. A straight mean of these results 
and those of \citet{bw92} then gives the following 
abundances for 47 Tuc, based on eight analyses of seven 
giants:  $<$[Fe/H]$_{\rm II}>$ = --0.70 and 
$<$[Fe/H]$_{\rm I}>$ = --0.65.

\subsubsection{NGC 7006; $E(B$--$V)$ = 0.10, (m--M)$_{0}$ = 18.00\label{Arevabunds.g3.n7006}}

     Of the six giants analyzed by \citet{kraft+98}, two 
are variable stars at or near the tip of the giant branch, 
and thus the colors and magnitudes measured by \citet{sw67}
might not have conformed to the properties of these stars 
at the time that the high-resolution observations were 
obtained. These two stars have therefore been omitted from 
the present study, reducing the number of giants available 
for analysis to four; thus, NGC 7006 has been placed in 
Group 3 even though the previous analysis is that of LTG. 

     Inspection of Table~\ref{table2} reveals that for 
these four stars the mean value of T$_{\rm eff}$ derived from 
the $($$B$--$V$$)_{0}$ colors (values of $V$--$K$ are unknown) differs
significantly from the mean value based on the Fe~{\sc i} 
excitation plot. This suggests that the assumed value 
of $E(B$--$V)$ = 0.05 (June 22, 1999 revision of Harris 
1996\nocite{harris96}) is too small. Agreement between the two 
methods yields $E(B$--$V)$ = 0.10 for this cluster. Earlier 
estimates of $E(B$--$V)$ range from 0.05 \citep{sz78} to 0.10 
\citep{claria+94}. The \citet{bh82} reddening maps suggest 
$E(B$--$V)$ = 0.09 and the \citet{schlegel+98} dust maps yield 
$E(B$--$V)$ = 0.08.  Thus our suggested reddening value is not 
outside the range of plausible values derived from 
non-spectroscopic considerations.

     In Table~\ref{tableA9} we list the original and 
revised values of T$_{\rm eff}$, log~$g$ and the revised 
Fe~{\sc i} and Fe~{\sc ii} abundances for the four 
non-variable giants.

Analyses of stars having T$_{\rm eff}$ $<$ 4000K are probably less 
reliable than those for hotter stars, so we give star I-1 half 
weight in calculating the means. The averages are 
$<$[Fe/H]$_{\rm II}>$ = --1.48 ($\sigma$ = 0.04) and 
$<$[Fe/H]$_{\rm I}>$ = --1.61 ($\sigma$ = 0.02). Thus the 
derived iron underabundance is --0.13, a value rather similar to 
what was found in M3 and M92.

\subsubsection{NGC 3201; $<$$E(B$--$V)$$>$ = 0.25, (m--M)$_{0}$ = 13.61\label{Arevabunds.g3.n3201}}

     In a recent extensive analysis of 18 giants in this cluster, 
\citet{gw98} found evidence for a range of metallicity in this
cluster, as well as a small range in reddening, with an average 
value near $E(B$--$V)$ = 0.25. They concluded that the existence of 
the metallicity range was uncertain and awaited further 
investigation. We proceed here on the assumption that the range 
is small and that a meaningful average value can be taken.       

     We deal here only with the 13 giants of their sample having 
measured values of $V$--$K$; of the remaining five giants, only 
rather uncertain values of $B$--$V$ are available and for these we 
are unable to estimate reliable values of T$_{\rm eff}$ based on the 
colors. We list in Table~\ref{tableA10} the values of T$_{\rm eff}$ 
and log~$g$ given by \citet{gw98}, based on the Fe~{\sc i} 
excitation plot and equivalence of [Fe/H] derived from 
Fe~{\sc i} and Fe~{\sc ii},\footnote{
The \citet{gw98} paper does not list separate values of log 
$\epsilon$(Fe~{\sc i}) and log $\epsilon$(Fe~{\sc ii}), but 
these do not differ by more than 0.01 dex \citep{gonzalez02}.} 
and also list the values of these quantities derived from 
$V$--$K$ corrected for the color excess, the \citet{alonso+99} 
T$_{\rm eff}$ scale, and the true distance modulus given above 
(June 22, 1999 revision of Harris 1996\nocite{harris96}, plus 
0.07 dex).

     The mean difference in the sense {\it T$_{\rm eff}$(rev) 
-- T$_{\rm eff}$(orig)} is +8K $\pm$ 14K ($\sigma$ = 50K), a 
small value, as expected, since values of T$_{\rm eff}$ derived 
from the Fe~{\sc i} excitation plot generally agree with 
those based on the \citet{alonso+99} T$_{\rm eff}$ scale. The 
mean difference in log~$g$, in the same sense, is +0.13 
$\pm$ 0.07 ($\sigma$ = 0.25), and thus we may expect that 
the value of $<$[Fe/H]$_{\rm II}>$ derived here will be 
about 0.05 dex higher than the value derived by 
\citet{gw98}, whereas $<$[Fe/H]$_{\rm I}>$ will be little 
changed.  For these 13 giants, we obtain 
$<$[Fe/H]$_{\rm II}>$ = --1.48 ($\sigma$ = 0.10), whereas 
the same 13 stars originally yielded $<$[Fe/H]$_{\rm II}>$ 
= $<$[Fe/H]$_{\rm I}>$ = --1.53 ($\sigma$ = 0.09).  The value 
of $<$[Fe/H]$_{\rm I}>$ remains unchanged.

     \citet{gw98} employed Kurucz models with convective
overshooting. Transforming the results to those that would
have been derived from MARCS models, we finally obtain 
$<$[Fe/H]$_{\rm II}>$ = --1.56 $\pm$ 0.10 and 
$<$[Fe/H]$_{\rm I}>$ = --1.54 $\pm$ 0.09.

\subsubsection{NGC 6752; $E(B$--$V)$ = 0.04, (m--M)$_{0}$ = 13.07\label{Arevabunds.g3.n6752}}

     The adopted value of $E(B$--$V)$ of 0.04 (June 22, 1999
revision of Harris 1996\nocite{harris96}) is in reasonable 
agreement with the \citet{schlegel+98} value of 0.055. From the 
literature, we find that \citet{ndc95} analyzed five giants. 
Comparing our derived values of T$_{\rm eff}$ and log~$g$ with 
theirs, we find mean offsets of $\Delta$T$_{\rm eff}$ = --15K 
($\sigma$ = 17K) and $\Delta$log~$g$ = --0.03 ($\sigma$ = 0.06) 
in the sense {\it this study minus Norris \& Da Costa}; 
within the errors, the parameters are identical. Norris \& Da 
Costa\nocite{ndc95} derived abundances from $\sim$90 to 100 
Fe~{\sc i} and 6 to 8 Fe~{\sc ii} lines. We made small 
adjustments ($\leq$ 0.04 dex) to their abundances to bring 
the log {\it gf}-values into agreement with those of 
Blackwell {\it et al.}  Based on their Kurucz models with 
convective overshooting, we find adjusted mean values of 
$<$[Fe/H]$_{\rm II}>$ = --1.42 ($\sigma$ = 0.04) and 
$<$[Fe/H]$_{\rm I}>$ = --1.50 ($\sigma$ = 0.04). On changing 
to MARCS models, these become $<$[Fe/H]$_{\rm II}>$ =
 --1.50 $\pm$ 0.04 and $<$[Fe/H]$_{\rm I}>$ = --1.51 $\pm$ 
0.04.

    Somewhat lower abundances are found from an adjusted 
analysis of three stars in a study by \citet{minniti+93}.  One 
star (A29) overlaps with \citet{ndc95} and provides a means of 
comparing EWs. From 22 common Fe~{\sc i} and Fe~{\sc ii} lines, 
we find $\Delta$EW in the sense {\it Minniti {\it et al.}\ 
minus Norris \& Da Costa} = +0.4 $\pm$ 1.5 ($\sigma$ = 7.0) 
m\AA; on average, the EW measurements are remarkably similar. 
\citet{minniti+93} measured four Fe~{\sc ii} lines in these 
giants and only three Fe~{\sc ii} lines in common with 
\citet{ndc95}, and one of these unfortunately has a smaller 
measured EW by 18m\AA, a change of 60 percent, quite the
largest difference of all the 22 common lines measured. 

Relative to the three Minniti stars, our adjustments lead to a 
mean value of T$_{\rm eff}$ that is 58K $\pm$ 31K lower and log~$g$ 
that is +0.03 $\pm$ 0.09 higher than the values adopted by 
\citet{minniti+93}, which are based on values of T$_{\rm eff}$ 
consistent with the Fe~{\sc i} excitation plot. In turn this 
suggests that the adopted reddening for this cluster is about 
0.04 too low in $E(B$--$V)$, but such a change would cause a
disagreement with the value derived from the dust maps.  
Although in this low-metallicity domain, an increase of 58K in 
T$_{\rm eff}$ increases Fe~{\sc i} by +0.10, it decreases 
Fe~{\sc ii}, our major concern, by only 0.01 dex.  If we then 
simply adopt the values of T$_{\rm eff}$ and log~$g$ given by 
Minniti {\it et al.}, and adjust their results based on Kurucz 
models to MARCS models, we find mean values 
$<$[Fe/H]$_{\rm II}>$ = --1.71 ($\sigma$ = 0.05) and 
$<$[Fe/H]$_{\rm I}>$ = --1.55 ($\sigma$ = 0.06). Since in 
comparison with the 6 to 8 lines measured by \citet{ndc95}, 
only four lines of Fe~{\sc ii} lines were actually measured by 
Minniti {\it et al.}, we give half weight to the results of 
Minniti {\it et al.}\ study and find grand means of 
$<$[Fe/H]$_{\rm II}>$ = --1.57 $\pm$ 0.10 and 
$<$[Fe/H]$_{\rm I}>$ = --1.53 $\pm$ 0.01, correcting Norris \&
Da Costa to the Minniti {\it et al.}\ T$_{\rm eff}$ scale, and 
adjusting all abundances to conform to those from MARCS models. 
The final abundance for Fe~{\sc ii} is clearly less well 
determined than is the case for the other clusters.

\subsubsection{NGC 2298; $E(B$--$V)$ = 0.16, (m--M)$_{0}$ = 15.17\label{Arevabunds.g3.n2298}}

    This cluster has a color excess somewhat above our 
comfortable excess limit of $E(B$--$V)$ = 0.10, but we include it 
in Group 3 because (1) it lies in an abundance regime between 
[Fe/H] = --1.5 and --2.2 which contains few clusters analyzed 
by LTG and (2) it has been subject to an abundance study in 
which values of T$_{\rm eff}$ were derived from colors but which 
were also compared with values derived from the Fe~{\sc i} 
excitation plot \citep{mcwilliam+92}. Three giants were 
analyzed.  Although many lines of Fe~{\sc i} were measured, 
only two lines of Fe~{\sc ii} were considered. We applied small 
adjustments to the adopted log {\it gf}-values to bring them 
into agreement with Blackwell {\it et al.}, as discussed 
earlier.  Since many Blackwell {\it et al.}\ and O'Brian log 
{\it gf}-values were adopted for the Fe~{\sc i} lines, we made 
no corresponding attempt to adjust the Fe~{\sc i} log 
{\it gf}-values. 

     McWilliam {\it et al.}\ found that their values of 
T$_{\rm eff}$ derived from $($$V$--$K$$)_{0}$ \citep{frogel+83} agreed with 
values derived from the Fe~{\sc i} excitation plot if $E(B$--$V)$ 
is taken to be 0.16; we therefore adopt that value of $E(B$--$V)$ 
here, although this value does not agree very well with that
derived from the Schlegel {\it et al}.\ dust maps, 
{\it i.e.}, $E(B$--$V)$ = 0.21. Using the Alonso {\it et al.}\ 
scale of T$_{\rm eff}$ versus $($$V$--$K$$)_{0}$, we derive values of 
T$_{\rm eff}$ that are only 20K smaller than those adopted by 
McWilliam {\it et al.}, and our values of log~$g$ average 
0.15 dex lower, based on the true distance modulus cited 
above (which corresponds to the Harris compilation for a value 
of $E(B$--$V)$ = 0.16 and the addition of +0.07 to the modulus). 

Our adjustment in T$_{\rm eff}$, log~$g$ and log {\it gf}-values 
lead then to $<$[Fe/H]$_{\rm II}>$ = --1.97 ($\sigma$ = 0.09) 
and $<$[Fe/H]$_{\rm I}>$ = --1.93 ($\sigma$ = 0.08).  No model 
adjustment was deemed necessary since McWilliam {\it et al.}\ 
also employed MARCS models.  These are our preferred estimates 
of Fe abundances, since they are based on the view that the 
Fe~{\sc i} excitation plot provides the correct estimate of 
T$_{\rm eff}$.  If, on the other hand, we had chosen $E(B$--$V)$ =
0.21, our T$_{\rm eff}$ estimates would have been hotter by 60K, and 
the corresponding values of [Fe/H] would have been 0.02 dex lower 
for Fe~{\sc ii} and 0.08 dex higher for Fe~{\sc i}.  The value of 
[Fe/H]$_{\rm II}$ for this cluster thus changes only slightly 
with the 0.05 mag change in $E(B$--$V)$.

\subsubsection{NGC 6397; $E(B$--$V)$ = 0.24, (m--M)$_{0}$ = 11.62\label{Arevabunds.g3.n6397}}

     This cluster also has $E(B$--$V)$ above our adopted cutoff 
of 0.10, but we include it because its metallicity is presumed 
to lie in the region between --1.5 and --2.2. In addition, we 
base our revised analysis on six giants for which 3 to 8 
Fe~{\sc ii} lines were measured. Thus we deal with four giants 
analyzed by Minniti {\it et al.}\ (1993;\nocite{minniti+93} 
Stars 302, 603, 669, A331) and two by Norris \& Da Costa 
(1995;\nocite{ndc95} Stars ROB 211 and ROB 469). The latter 
two stars have observed values of $V$--$K$ \citep{frogel+83}, 
as do two of the four giants of Minniti {\it et al.}  There is 
no overlap between the stars of these two investigations, so 
EWs cannot be directly compared.  A third study by 
CG97\nocite{cg97} includes ROB 211 among its three giants.  

Except for one line, the CG97 EWs for ROB 211 exceed those of 
Norris \& Da Costa.  Below 50 m\AA, the difference amounts to 30 
to 50 percent. The situation is quite serious for the five 
lines of Fe~{\sc ii}, and probably explains why the Fe~{\sc ii} 
abundance of CG97 is about 0.2 dex larger than that of Norris 
\& Da Costa. The only EW comparison between Minniti 
{\it et al.}\ and Norris \& Da Costa is that of star A29 in NGC 
6752, already discussed, and in that case we found excellent 
agreement between the two investigations, except for one 
discordant Fe~{\sc ii} line. We therefore tentatively drop the 
CG97 study of ROB 211 from further consideration, recognizing 
that evidence for doing so is not entirely compelling.

    Norris \& Da Costa measured 11 Fe~{\sc ii} lines; Minniti 
{\it et al.}\ measured four. The T$_{\rm eff}$'s we derive based 
on $($$V$--$K$$)_{0}$ or $($$B$--$V$$)_{0}$ and the Alonso {\it et al.}\ 
T$_{\rm eff}$ scale run on the average 70K lower than those of 
Norris \& Da Costa and 90K lower than those of Minniti 
{\it et al.}, the latter having been adjusted to satisfy the 
Fe~{\sc i} excitation plot. This latter result suggests that 
$E(B$--$V)$ for NGC 6397 should be increased from 0.18 to 0.24, 
even though the Schlegel {\it et al.}\ dust maps yield 
$E(B$--$V)$ = 0.185, close to the value tabulated by Harris. 

Following the procedures developed in the present study, and 
adopting our photometrically calibrated T$_{\rm eff}$'s increased 
by 90K and adjusting the log~$g$'s accordingly, we derive 
$<$[Fe/H]$_{\rm II}>$ = --2.04 $\pm$ 0.02 and 
$<$[Fe/H]$_{\rm I}>$ = --1.83 $\pm$ 0.02, based on the 
Minniti {\it et al.}\ abundances and Kurucz models. The 
corresponding treatment of the Norris \& Da Costa data yields 
$<$[Fe/H]$_{\rm II}>$ = --1.91 $\pm$ 0.07 and
$<$[Fe/H]$_{\rm I}>$ = --2.02 $\pm$ 0.02. Taking means and 
adjusting to MARCS models leads to 
$<$[Fe/H]$_{\rm II}>$ = --2.02 $\pm$ 0.07  and 
$<$[Fe/H]$_{\rm I}>$ = --2.04 $\pm$ 0.03.  As was the case in 
our discussion of NGC 2298, adoption of the (in this case) 
lower reddening would have only a slight effect on the derived
value of $<$[Fe/H]$_{\rm II}>$.
 
      In a more recent study of NGC 6397, \citet{castilho+00}
analyzed 16 stars of which 5 were giants and the balance were 
subgiants having M$_{bol}$ $\geq$ +0.4. They assumed $E(B$--$V)$ 
= 0.18, and assigned values of T$_{\rm eff}$ from 
reddening-corrected colors, based on various scales published 
prior to those of \citet{alonso+99} and \citet{houdashelt+00}.
Values of log~$g$ were derived from the known distance 
modulus, estimated T$_{\rm eff}$'s, and assumed masses of 
0.8~$M_{\sun}$, and checked with the demand that 
[Fe/H]$_{\rm I}$ should equal [Fe/H]$_{\rm II}$. They found 
that giants and subgiants yielded the same value of [Fe/H].

      In attempting to make adjustments to the methods employed 
here, we considered only the five giants, confining attention 
to stars with values of T$_{\rm eff}$ and log~$g$ similar to those 
studied in the other clusters.  However, since 
\citet{castilho+00} did not publish linelists for Fe~{\sc i} 
and Fe~{\sc ii}, we are unable to check the value of T$_{\rm eff}$ 
that could be derived from the Fe~{\sc i} excitation plot, and
use this to estimate the correct choice of $E(B$--$V)$. However, 
if we assume that [Fe/H]$_{\rm I}$ was set essentially equal 
to [Fe/H]$_{\rm II}$, as indirectly indicated by these authors, 
we can estimate the revisions to the Fe abundances, based on 
procedures adopted here.

\citet{castilho+00} adopted MARCS models, and employed log 
{\it gf}-values from \citet{weise+69} which in turn are close 
to the system adopted by \citet{lambert+96}. For Fe~{\sc ii}, 
we find [Fe/H]$_{\rm II}$ = --2.01 ($\sigma$ = 0.08), 
independent of whether $E(B$--$V)$ = 0.18 or 0.24 (the small 
shift in log~$g$ tends to compensate for the T$_{\rm eff}$ change 
in the case of Fe~{\sc ii}). The effect on Fe derived from 
Fe~{\sc i} is more serious: for $E(B$--$V)$ = 0.24, we find 
[Fe/H]$_{\rm I}$ = --1.88 $\pm$ 0.08 whereas for $E(B$--$V)$ = 
0.18, we find [Fe/H]$_{\rm I}$ = --2.09 $\pm$ 0.06.  
Evidently these 5 giants yield the same value of 
[Fe/H]$_{\rm II}$ as we found earlier, after adjusting the 
Minniti {\it et al.}\ and Norris \& Da Costa values.

%




\clearpage

\clearpage

\clearpage

\clearpage

\clearpage

\begin{deluxetable}{ccccc}
\tablenum{1}
\tablecaption{log {\it gf}-values for Fe~{\sc ii} lines studied by LTG\label{table1}}
\tablewidth{0pt}
\tablehead{
\colhead{$\lambda$}    &
\colhead{log {\it gf}} &
\colhead{log {\it gf}} &
\colhead{log {\it gf}} &
\colhead{log {\it gf}} \\
\colhead{\AA }          &
\colhead{LTG}           &
\colhead{Lambert\tablenotemark{(a)}} &
\colhead{Biemont\tablenotemark{(a)}} &
\colhead{Blackwell\tablenotemark{(a)}}                              
}                      
\startdata
6149.25 &  --2.72 & --2.73 & --2.70 & --2.75 \\
6369.46 &  --4.25 & --4.20 & --4.21 & --4.19 \\
6416.93 &  --2.79 & --2.69 & --2.70 & --2.68 \\
6432.68 &  --3.71 & --3.59 & --3.60 & --3.57 \\
6456.39 &  --2.08 & \nodata& \nodata& --2.13 \\
6516.08 &  --3.45 & \nodata& \nodata& --3.28 \\
\enddata
\tablenotetext{(a)}{The log {\it gf}-values of Lambert {\it et al.},
Biemont {\it et al.}, and Blackwell {\it et al.}\ have been adjusted 
to conform to an adopted solar Fe abundance of 7.52.}
\end{deluxetable}

\clearpage

\begin{deluxetable}{rccccrcl}
\tabletypesize{\footnotesize}
\tablenum{2}
\tablecaption{Parameters for 11 Clusters Analyzed by LTG\label{table2}}
\tablewidth{0pt}
\tablehead{
\colhead{Cluster}                              & 
\colhead{$E(B$--$V)$}                             &
\colhead{$E(B$--$V)$}                             &
\colhead{$E(B$--$V)$\tablenotemark{(a)}}          &
\colhead{(m--M)$_{0}$\tablenotemark{(b)}}       &
\colhead{$\Delta$T$_{\rm eff}$\tablenotemark{(c)}} &
\colhead{No.}                                  &
\colhead{Ref.}                                 \\
\colhead{}                                     &
\colhead{Harris}                               &
\colhead{Schlegel}                             &
\colhead{adopt}                                &
\colhead{adopt}                                &
\colhead{IRFM--Fe~{\sc i}$_{exc}$}             &
\colhead{stars\tablenotemark{(d)}}             &
\colhead{}
}                      
\startdata
 N288 & 0.03 & 0.01 & 0.03~ & 14.66 &    +4K$\pm$19K & 10 &  Shetrone \& Keane 2000 \\
 N362 & 0.05 & 0.04 & 0.05~ & 14.70 &    +4K$\pm$16K & ~9 &  Shetrone \& Keane 2000 \\
   M3 & 0.01 & 0.01 & 0.01~ & 15.02 &    +6K$\pm$13K & 17 &  Kraft {\it et al.}\ 1999 \\
   M5 & 0.03 & 0.04 & 0.03~ & 14.42 &   +21K$\pm$~9K & 25 &  Ivans {\it et al.}\ 2001 \\
   M4 & 0.36 & 0.50 & 0.33v & 11.61 &   +21K$\pm$58K & 24 &  Ivans {\it et al.}\ 1999 \\
  M13 & 0.02 & 0.02 & 0.02~ & 14.42 &  --39K$\pm$10K & 28 &  Kraft {\it et al.}\ 1997 \\
  M10 & 0.28 & 0.29 & 0.24~ & 13.41 &   +88K$\pm$22K & ~9 &  Kraft {\it et al.}\ 1995 \\
  M92 & 0.02 & 0.02 & 0.02~ & 14.75 &  --10K$\pm$14K & ~6 &  Shetrone 1996 \\
      &      &      &       &       &                &    &  Langer {\it et al.}\ 1999 \\
  M71 & 0.25 & 0.31 & 0.32~ & 12.83 &  --60K$\pm$~8K & 10 &  Sneden {\it et al.}\ 1994 \\
N7006 & 0.05 & 0.08 & 0.10~ & 18.00 & --102K$\pm$16K & ~4 &  Kraft {\it et al.}\ 1998 \\
  M15 & 0.10 & 0.11 & 0.10~ & 15.25 &   +31K$\pm$32K & ~7 &  Sneden {\it et al.}\ 1997 \\
\enddata
\tablenotetext{(a)}{Adopted color excess, assuming $\Delta$T$_{\rm eff}$ 
$\sim$0K (see text).}
\tablenotetext{(b)}{True distance modulus corrected for A$_{V}$ = 
3.1$\times E(B$--$V)$, where the color excess is that of column 3.}
\tablenotetext{(c)}{Difference between T$_{\rm eff}$ based on colors, 
employing the \citet{alonso+99} calibration and $E(B$--$V)_{(Harris)}$, 
and T$_{\rm eff}$ based on the Fe~{\sc i} excitation plot. The quoted
errors are $\sigma$/$\sqrt{N}$. The mean $\Delta$T$_{\rm eff}$ 
for the seven clusters having $\Delta$T$_{\rm eff}$'s $<$ $|40|$K 
and $E(B$--$V)$ $\leq$ 0.10 is +2K $\pm$ 20K.} 
\tablenotetext{(d)}{Number of stars analyzed for this compilation. 
Total number is 149.}
\end{deluxetable}

\clearpage

\begin{deluxetable}{ccccccl}
\tablenum{3}
\tablecaption{True Distance Moduli of Five Key Clusters\label{table3}}
\tablewidth{0pt}
\tablehead{
\colhead{Cluster}           & 
\colhead{[Fe/H]$_{\rm II}$} &
\colhead{(m--M)$_{0}$}      &
\colhead{(m--M)$_{0}$}      &
\colhead{(m--M)$_{0}$}      &
\colhead{(m--M)$_{0}$}      &
\colhead{Remark}            \\
\colhead{}                  &
\colhead{this study}        &
\colhead{C2000}             &
\colhead{R1997}             &
\colhead{LTG}               &
\colhead{this study}        &
\colhead{}
}                      
\startdata
M15 & --2.42  &\nodata & 15.25 & 15.26 & 15.25 & Reid $E(B$--$V)$=0.07 \\
    &         &        & 15.51 &       &       & Reid $E(B$--$V)$=0.11  \\
M92 & --2.38  &  14.71 & 14.93 & 14.49 & 14.75 &  \\
M13 & --1.60  &  14.40 & 14.48 & 14.08 & 14.42 &  LTG K1997 \\
    &         &        &       &       & 14.26 &  LTG P1996 \\
M3  & --1.50  &\nodata &\nodata& 14.79 & 15.02 &  \\
M5  & --1.26  &  14.59 & 14.45 & 14.03 & 14.42 &  LTG S1992 \\
    &         &        &       &       & 14.40 &  LTG I2001 \\
\enddata
\tablerefs{
C2000 $\equiv$ Carretta {\it et al.}\ 2000\nocite{carretta+00};
I2001 $\equiv$ Ivans {\it et al.}\ 2001\nocite{ivans+01}.
K1997 $\equiv$ Kraft {\it et al.}\ 1997\nocite{kraft+97};
P1996 $\equiv$ Pilachowski {\it et al.}\ 1996\nocite{pilachowski+96};
R1997 $\equiv$ Reid 1997\nocite{reid97};
S1992 $\equiv$ Sneden {\it et al.}\ 1992\nocite{sneden+92}
}
\end{deluxetable}

\clearpage

\begin{deluxetable}{ccrcrccccr}
\tabletypesize{\footnotesize}
\tablenum{4}
\tablecaption{Globular Cluster Fe~{\sc ii} \& Fe~{\sc i} Abundances Segregated by Model\label{table4}}
\tablewidth{0pt}
\tablehead{
\colhead{Cluster}           &
\colhead{[Fe/H]$_{\rm II}$} &
\colhead{$\sigma$}          &
\colhead{[Fe/H]$_{\rm I}$}  &
\colhead{$\sigma$}          &
\colhead{[Fe/H]$_{\rm II}$} &
\colhead{[Fe/H]$_{\rm I}$}  &
\colhead{[Fe/H]$_{\rm II}$} &
\colhead{[Fe/H]$_{\rm I}$}  &
\colhead{$<$$W'$$>$}            \\
\colhead{}                  &
\colhead{MARCS}             &
\colhead{}                  &
\colhead{MARCS}             &
\colhead{}                  &
\colhead{Kurucz}            &
\colhead{Kurucz}            &
\colhead{Kurucz}            &
\colhead{Kurucz}            &
\colhead{}                  \\
\colhead{}                  &
\colhead{}                  &
\colhead{}                  &
\colhead{}                  &
\colhead{}                  &
\colhead{on}                &
\colhead{on}                &
\colhead{off}               &
\colhead{off}               &
\colhead{}                  
}                      
\startdata
 	& 	& 	& 	& 	Group 1 & & & & & \\
 M5      & --1.26 & 0.06 & --1.36 & 0.07 & --1.19 & --1.36 & --1.19 & --1.37 & 3.73 \\
NGC 362  & --1.34 & 0.07 & --1.31 & 0.03 & --1.27 & --1.31 & --1.27 & --1.32 & 3.72 \\
NGC 288  & --1.41 & 0.04 & --1.36 & 0.07 & --1.33 & --1.35 & --1.35 & --1.36 & 3.66 \\
 M3      & --1.50 & 0.03 & --1.58 & 0.06 & --1.42 & --1.57 & --1.43 & --1.59 & 3.25\tablenotemark{(a)} \\
 M13     & --1.60 & 0.08 & --1.63 & 0.06 & --1.52 & --1.62 & --1.53 & --1.64 & 3.27 \\
 M92     & --2.38 & 0.07 & --2.50 & 0.12 & --2.32 & --2.40 & --2.38 & --2.48 & 1.82\tablenotemark{(b)}  \\
 M15     & --2.42 & 0.07 & --2.50 & 0.11 & --2.36 & --2.40 & --2.42 & --2.48 & 1.56 \\
 	& 	& 	& 	& 	Group 2 & & & & & \\
 M71     & --0.81 & 0.07 & --0.82 & 0.05 & --0.74 & --0.75 & --0.74 & --0.75 & 4.64 \\
 M4      & --1.15 & 0.08 & --1.15 & 0.05 & --1.08 & --1.15 & --1.08 & --1.09 & 3.87 \\
 M10     & --1.51 & 0.09 & --1.50 & 0.05 & --1.43 & --1.49 & --1.44 & --1.51 & 3.42 \\
 	& 	& 	& 	& 	Group 3 & & & & & \\
 47 Tuc  & --0.70 & 0.09 & --0.65 & 0.04 & --0.63 & --0.58 & --0.63 & --0.58 & 4.53 \\
NGC 7006 & --1.48 & 0.04 & --1.61 & 0.02 & --1.40 & --1.60 & --1.41 & --1.62 & 3.55\tablenotemark{(a)} \\
NGC 3201 & --1.56 & 0.10 & --1.54 & 0.09 & --1.48 & --1.53 & --1.49 & --1.55 & 3.42 \\
NGC 6752 & --1.57 & 0.10 & --1.53 & 0.02 & --1.49 & --1.52 & --1.50 & --1.54 & 3.41 \\
NGC 2298 & --1.97 & 0.09 & --1.93 & 0.08 & --1.91 & --1.83 & --1.97 & --1.91 & 2.27 \\
NGC 6397 & --2.02 & 0.07 & --2.04 & 0.03 & --1.96 & --1.94 & --2.02 & --1.86 & 2.17 \\
\enddata
\tablenotetext{(a)}{$W'$ value from our correlation of $W'$ with 
$Q39$ taken from \citet{zinn80}.}
\tablenotetext{(b)}{Extrapolated value of $W'$ (based on values from
Rutledge {\it et al.}\ 1997a\nocite{rutledge+97a},b\nocite{rutledge+97b}).}
\end{deluxetable}

\clearpage

\begin{deluxetable}{crrrr}
\tablenum{5}
\tablecaption{Mean Differences in T$_{\rm eff}$: Houdashelt minus Alonso\label{table5}}
\tablewidth{0pt}
\tablehead{
\colhead{Cluster}                   &
\colhead{[Fe/H]$_{\rm II}$}         &
\colhead{$\Delta$T$_{\rm eff}$}         &
\colhead{$\Delta$[Fe/H]$_{\rm II}$} &
\colhead{$\Delta$[Fe/H]$_{\rm I}$}  \\
\colhead{}                  &
\colhead{MARCS}             &
\colhead{approx}            &
\colhead{}                  &
\colhead{}                  
}                      
\startdata
        M71 & --0.81 &   30K & --0.04 &  +0.02 \\
        M5  & --1.26 &   45K & --0.05 &  +0.04 \\
        M13 & --1.60 &   45K & --0.05 &  +0.05 \\
        M92 & --2.38 &  130K & --0.03 &  +0.15 \\
\enddata
\end{deluxetable}

\clearpage

\begin{deluxetable}{cccc}
\tablenum{6}
\tablecaption{Observed and Predicted Overionization of Fe ([Fe/H]$_{\rm I}$ - [Fe/H]$_{\rm II}$)\label{table6}}
\tablewidth{0pt}
\tablehead{
\colhead{Cluster}           &
\colhead{[Fe/H]$_{\rm II}$} &
\colhead{$\Delta$[Fe/H]}    &
\colhead{$\Delta$[Fe/H]}    \\
\colhead{}                  &
\colhead{MARCS}             &
\colhead{observed}          &
\colhead{pred TI99}         
}                      
\startdata
        M71 & --0.81 &  --0.01 &  --0.05 \\
        M4  & --1.15 &   ~0.00 &  --0.15 \\
        M5  & --1.26 &  --0.10 &  --0.18 \\
       N362 & --1.34 &   +0.03 &  --0.20 \\
       N288 & --1.40 &   +0.05 &  --0.20 \\
        M3  & --1.50 &  --0.08 &  --0.22 \\
        M10 & --1.51 &   +0.01 &  --0.22 \\
        M13 & --1.60 &  --0.03 &  --0.23 \\
        M92 & --2.38 &  --0.12 &  --0.28 \\
        M15 & --2.42 &  --0.08 &  --0.28 \\
\enddata
\end{deluxetable}

\clearpage

\begin{deluxetable}{rrcccc}
\tablenum{7}
\tabletypesize{\footnotesize}
\tablecaption{Compilation of Cluster [Fe/H]$_{\rm II}$ Estimates\label{table7}}
\tablewidth{0pt}
\tablehead{
\colhead{NGC}               &
\colhead{Alt.\ Name}        &
\colhead{ZW84}              &
\colhead{MARCS}             &
\colhead{Kurucz-on}         &
\colhead{Kurucz-off}     
}                      
\startdata
104  & 47Tuc   & -0.71  & -0.88  & -0.79  & -0.78 \\
288  &         & -1.40  & -1.34  & -1.26  & -1.27 \\
362  &         & -1.27  & -1.27  & -1.20  & -1.21 \\
1261 &         & -1.29  & -1.26  & -1.19  & -1.19 \\
     & Eridanus& -1.35  & -1.42  & -1.34  & -1.35 \\
1851 &         & -1.33  & -1.19  & -1.11  & -1.12 \\
1904 &         & -1.68  & -1.64  & -1.57  & -1.59 \\
2298 &         & -1.81  & -2.07  & -2.00  & -2.04 \\
2808 &         & -1.37  & -1.29  & -1.22  & -1.23 \\
     & Pal 3   & -1.78  & -1.66  & -1.59  & -1.61 \\
3201 &         & -1.56  & -1.46  & -1.39  & -1.40 \\
     & Pal 4   & -2.20  & -1.43  & -1.35  & -1.37 \\
4147 &         & -1.80  & -1.79  & -1.72  & -1.75 \\
4372 &         & -2.08  & -2.29  & -2.23  & -2.28 \\
4590 & M68     & -2.09  & -2.43  & -2.37  & -2.43 \\
4833 &         & -1.86  & -2.06  & -2.00  & -2.04 \\
5053 &         & -2.58  & -2.41  & -2.35  & -2.41 \\
5286 &         & -1.79  & -1.69  & -1.62  & -1.65 \\
5694 &         & -1.92  & -2.10  & -2.04  & -2.08 \\
5897 &         & -1.68  & -2.09  & -2.02  & -2.07 \\
5904 & M5      & -1.40  & -1.32  & -1.24  & -1.25 \\
5927 &         & -0.30  & -0.67  &\nodata &\nodata\\
5986 &         & -1.67  & -1.61  & -1.54  & -1.56 \\
     & Pal 14  & -1.47  & -1.61  & -1.54  & -1.56 \\
6093 & M80     & -1.68  & -1.76  & -1.69  & -1.72 \\
6101 &         & -1.81  & -2.13  & -2.07  & -2.12 \\
6121 & M4      & -1.28  & -1.22  & -1.15  & -1.15 \\
6144 &         & -1.75  & -2.10  & -2.03  & -2.08 \\
6171 & M107    & -0.99  & -1.10  & -1.02  & -1.02 \\
6205 & M13     & -1.65  & -1.58  & -1.51  & -1.53 \\
6218 & M12     & -1.61  & -1.34  & -1.25  & -1.26 \\
6235 &         & -1.40  & -1.39  & -1.31  & -1.32 \\
6254 & M10     & -1.60  & -1.48  & -1.41  & -1.43 \\
6266 & M62     & -1.29  & -1.19  & -1.11  & -1.12 \\
6273 & M19     & -1.68  & -1.84  & -1.77  & -1.79 \\
6304 &         & -0.59  & -0.73  &\nodata &\nodata\\
6352 &         & -0.51  & -0.78  & -0.70  & -0.69 \\
6366 &         & -0.99  & -0.82  & -0.74  & -0.73 \\
6362 &         & -1.08  & -1.15  & -1.07  & -1.07 \\
6397 &         & -1.91  & -2.12  & -2.06  & -2.11 \\
6496 &         & -0.48  & -0.78  & -0.70  & -0.69 \\
6522 &         & -1.44  & -1.42  & -1.35  & -1.36 \\
6535 &         & -1.75  & -1.80  & -1.74  & -1.76 \\
6544 &         & -1.56  & -1.41  & -1.34  & -1.35 \\
6541 &         & -1.83  & -1.83  & -1.76  & -1.78 \\
6624 &         & -0.35  & -0.78  & -0.70  & -0.69 \\
6626 &         & -1.44  & -1.19  & -1.11  & -1.12 \\
6638 &         & -1.15  & -1.04  & -0.96  & -0.92 \\
6637 & M69     & -0.59  & -0.89  & -0.81  & -0.80 \\
6681 & M70     & -1.51  & -1.60  & -1.52  & -1.54 \\
6712 &         & -1.01  & -1.10  & -1.02  & -1.02 \\
6715 & M54     & -1.43  & -1.47  & -1.40  & -1.41 \\
6717 & Pal 9   & -1.32  & -1.27  & -1.20  & -1.21 \\
6723 &         & -1.09  & -1.11  & -1.03  & -1.03 \\
6752 &         & -1.54  & -1.46  & -1.39  & -1.41 \\
6809 & M55     & -1.82  & -1.85  & -1.78  & -1.82 \\
     & Pal 11  & -0.70  & -0.80  &\nodata &\nodata\\
6838 & M71     & -0.58  & -0.82  & -0.74  & -0.73 \\
6981 & M72     & -1.54  & -1.42  & -1.35  & -1.36 \\
7078 & M15     & -2.15  & -2.45  & -2.39  & -2.46 \\
7089 & M2      & -1.62  & -1.56  & -1.49  & -1.51 \\
7099 & M30     & -2.13  & -2.33  & -2.26  & -2.32 \\
     & Pal 12  & -1.14  & -0.95  & -0.87  & -0.86 \\
7492 &         & -1.51  & -1.69  & -1.62  & -1.64 \\
\enddata
\end{deluxetable}

\clearpage

\begin{deluxetable}{crrrrrrl}
\tablenum{A1}
\tablecaption{Fe Abundances for Seven Giants in NGC~362\label{tableA1}}
\tablewidth{0pt}
\tablehead{
\colhead{Star}              & 
\colhead{T$_{\rm eff}$}         &
\colhead{log~$g$}           &
\colhead{T$_{\rm eff}$}         &
\colhead{log~$g$}           &
\colhead{[Fe/H]$_{\rm II}$} &
\colhead{[Fe/H]$_{\rm I}$}  &
\colhead{Remark}            \\
\colhead{}                  &
\colhead{orig}              &
\colhead{orig}              &
\colhead{rev}               &
\colhead{rev}               &
\colhead{rev}               &
\colhead{rev}               &
\colhead{}
}                      
\startdata
  1423   &  3950  &  0.10 &   4000 &  0.48  &   --1.26  &     --1.31 & \\
  1334   &  3975  &  0.40 &   4000 &  0.40  &   --1.34  &     --1.30 & \\
  2423   &  4000  &  0.40 &   4025 &  0.58  &   --1.32  &     --1.30 & \\
  1137   &  4000  &  0.70 &   4050 &  0.64  &   --1.44  &     --1.35 & \\
    77   &  4075  &  0.20 &   4150 &  0.55  &   --1.26  &     --1.26 & AGB? \\
  2127   &  4110  &  0.60 &   4050 &  0.66  &   --1.35  &     --1.35 & \\
  1159   &  4125  &  0.80 &   4100 &  0.59  &   --1.42  &     --1.33 & \\
\enddata
\end{deluxetable}

\clearpage

\begin{deluxetable}{crrrrrr}
\tablenum{A2}
\tablecaption{Fe Abundances for Eight Giants in NGC~288\label{tableA2}}
\tablewidth{0pt}
\tablehead{
\colhead{Star}              & 
\colhead{T$_{\rm eff}$}         &
\colhead{log~$g$}           &
\colhead{T$_{\rm eff}$}         &
\colhead{log~$g$}           &
\colhead{[Fe/H]$_{\rm II}$} &
\colhead{[Fe/H]$_{\rm I}$}  \\
\colhead{}                  &
\colhead{orig}              &
\colhead{orig}              &
\colhead{rev}               &
\colhead{rev}               &
\colhead{rev}               &
\colhead{rev}               
}                      
\startdata
   274 &    4025 &  0.70  &   4050 &   0.67 &    --1.47  &    --1.38  \\
   20c &    4050 &  0.60  &   4025 &   0.70 &    --1.43  &    --1.47  \\
   281 &    4125 &  0.60  &   4200 &   0.88 &    --1.40  &    --1.41  \\
   344 &    4180 &  0.80  &   4250 &   1.00 &    --1.34  &    --1.30  \\
   245 &    4250 &  0.80  &   4300 &   1.11 &    --1.36  &    --1.25  \\
   231 &    4300 &  1.10  &   4325 &   1.21 &    --1.40  &    --1.40  \\
   338 &    4325 &  1.30  &   4375 &   1.25 &    --1.45  &    --1.34  \\
   351 &    4330 &  1.20  &   4375 &   1.22 &    --1.41  &    --1.32  \\
\enddata
\end{deluxetable}

\clearpage

\begin{deluxetable}{crrrrrr}
\tablenum{A3}
\tablecaption{Fe Abundances for Four Giants in M3\label{tableA3}}
\tablewidth{0pt}
\tablehead{
\colhead{Star}              & 
\colhead{T$_{\rm eff}$}         &
\colhead{log~$g$}           &
\colhead{T$_{\rm eff}$}         &
\colhead{log~$g$}           &
\colhead{[Fe/H]$_{\rm II}$} &
\colhead{[Fe/H]$_{\rm I}$}  \\
\colhead{}                  &
\colhead{orig}              &
\colhead{orig}              &
\colhead{rev}               &
\colhead{rev}               &
\colhead{rev}               &
\colhead{rev}               
}                      
\startdata
   I-21  &  4200 &  0.75  &   4175 &   0.72  &    --1.47  &    --1.54  \\
  IV-77  &  4250 &  0.75  &   4300 &   0.88  &    --1.48  &    --1.54  \\
  IV-101 &  4200 &  0.85  &   4200 &   0.75  &    --1.48  &    --1.52  \\
  VZ 729 &  4200 &  0.85  &   4200 &   0.82  &    --1.45  &    --1.50  \\
\enddata
\end{deluxetable}

\clearpage

\begin{deluxetable}{crrrrrrl}
\tablenum{A4}
\tablecaption{Fe Abundances for 28 Giants in M13\label{tableA4}}
\tablewidth{0pt}
\tablehead{
\colhead{Star}              & 
\colhead{T$_{\rm eff}$}         &
\colhead{log~$g$}           &
\colhead{T$_{\rm eff}$}         &
\colhead{log~$g$}           &
\colhead{[Fe/H]$_{\rm II}$} &
\colhead{[Fe/H]$_{\rm I}$}  &
\colhead{Remark}            \\
\colhead{}                  &
\colhead{orig}              &
\colhead{orig}              &
\colhead{rev}               &
\colhead{rev}               &
\colhead{rev}               &
\colhead{rev}               &
\colhead{}
}                      
\startdata
 L 598  &  3900   &  0.00   &  3900  &   0.30    &    --1.53    &  --1.64 & \\
 L 629  &  3950   &  0.20   &  4010  &   0.36    &    --1.65    &  --1.65 & \\
 I-48   &  3920   &  0.30   &  3950  &   0.34    &    --1.61    &  --1.59 & \\
II-67   &  3950   &  0.20   &  3900  &   0.37    &    --1.40    &  --1.65 & \\
IV-25   &  4000   &  0.15   &  3975  &   0.38    &    --1.41    &  --1.66 & \\
II-90   &  4000   &  0.30   &  3960  &   0.38    &    --1.64    &  --1.65 & \\
 L 835  &  4090   &  0.55   &  4035  &   0.45    &    --1.52    &  --1.62 & \\
III-56  &  4100   &  0.65   &  4000  &   0.42    &    --1.57    &  --1.66 & \\
 L 853  &  4180   &  0.80   &  4030  &   0.48    &    --1.62    &  --1.67 & \\
 L 261  &  4230   &  0.85   &  4150  &   0.60    &    --1.68    &  --1.57 & \\
 L 262  &  4180   &  0.80   &  4125  &   0.60    &    --1.58    &  --1.58 & \\
II-34   &  4190   &  0.85   &  4140  &   0.63    &    --1.69    &  --1.66 & \\
III-73  &  4300   &  0.85   &  4210  &   0.69    &    --1.61    &  --1.59 & \\
II-76   &  4350   &  1.00   &  4250  &   0.80    &    --1.55    &  --1.67 & \\
I-13    &  4290   &  1.00   &  4230  &   0.80    &    --1.54    &  --1.57 & \\
III-59  &  4360   &  1.10   &  4290  &   0.85    &    --1.60    &  --1.53 & \\
III-52  &  4335   &  1.00   &  4275  &   0.87    &    --1.53    &  --1.65 & \\
 J3     &  4575   &  1.65   &  4520  &   1.31    &    --1.67    &  --1.65 & \\
 A1     &  4550   &  1.50   &  4550  &   1.34    &    --1.72    &  --1.74 & \\
I-12    &  4600   &  1.50   &  4610  &   1.46    &    --1.65    &  --1.60 & \\
IV-19   &  4650   &  1.50   &  4610  &   1.52    &    --1.61    &  --1.64 & \\
IV-22   &  4700   &  1.90   &  4750  &   2.12    &    --1.51    &  --1.52 & \\
II-9    &  4700   &  1.70   &  4680  &   1.75    &    --1.66    &  --1.62 & \\
II-41   &  4750   &  2.00   &  4650  &   1.62    &    --1.64    &  --1.56 & \\
I-72    &  4850   &  1.90   &  4825  &   1.97    &    --1.62    &  --1.69 & \\
II-28   &  4850   &  2.00   &  4750  &   1.57    &    --1.71    &  --1.79 & \\
II-1    &  4850   &  2.10   &  4850  &   2.05    &    --1.67    &  --1.60 & \\
I-54    &  4975   &  1.70   &  5050  &   1.74    &    --1.67    &  --1.66 & AGB \\
\enddata
\end{deluxetable}

\clearpage

\begin{deluxetable}{crrrrrrl}
\tablenum{A5}
\tablecaption{Fe Abundances for Six Giants in M92\label{tableA5}}
\tablewidth{0pt}
\tablehead{
\colhead{Star}              & 
\colhead{T$_{\rm eff}$}         &
\colhead{log~$g$}           &
\colhead{T$_{\rm eff}$}         &
\colhead{log~$g$}           &
\colhead{[Fe/H]$_{\rm II}$} &
\colhead{[Fe/H]$_{\rm I}$}  &
\colhead{Remark}            \\
\colhead{}                  &
\colhead{orig}              &
\colhead{orig}              &
\colhead{rev}               &
\colhead{rev}               &
\colhead{rev}               &
\colhead{rev}               &
\colhead{}
}                      
\startdata
 XII-8 &   4515 &   1.10 &     4475 &    0.93  &   --2.46 & --2.61 & \\
 V-45  &   4490 &   1.10 &     4450 &    0.94  &   --2.47 & --2.67 & \\
 XI-19 &   4515 &   1.10 &     4480 &    0.98  &   --2.30 & --2.43 & \\
 X-49  &   4180 &   0.10 &     4225 &    0.53  &   --2.31 & --2.43 & AGB??  \\
 B-19  &   4300 &   0.75 &     4330 &    0.65  &   --2.36 & --2.32 & \\
 B-95  &   4270 &   0.70 &     4250 &    0.59  &   --2.40 & --2.51 & \\
\enddata
\end{deluxetable}

\clearpage

\begin{deluxetable}{crrrrrr}
\tablenum{A6}
\tablecaption{Fe Abundances for Nine Giants in M15\label{tableA6}}
\tablewidth{0pt}
\tablehead{
\colhead{Star}              & 
\colhead{T$_{\rm eff}$}         &
\colhead{log~$g$}           &
\colhead{T$_{\rm eff}$}         &
\colhead{log~$g$}           &
\colhead{[Fe/H]$_{\rm II}$} &
\colhead{[Fe/H]$_{\rm I}$}  \\
\colhead{}                  &
\colhead{orig}              &
\colhead{orig}              &
\colhead{rev}               &
\colhead{rev}               &
\colhead{rev}               &
\colhead{rev}               
}                      
\startdata
 IV-38 & 4275 & 0.65 & 4125 & 0.50 & --2.42 & --2.59 \\
  S-4  & 4275 & 0.65 & 4160 & 0.53 & --2.51 & --2.59 \\
  S-1  & 4375 & 0.50 & 4280 & 0.55 & --2.45 & --2.62 \\
 II-64 & 4400 & 0.65 & 4470 & 1.00 & --2.37 & --2.40 \\
  S-3  & 4400 & 0.65 & 4520 & 1.02 & --2.33 & --2.30 \\
 II-75 & 4425 & 0.75 & 4300 & 0.66 & --2.40 & --2.55 \\
  S-7  & 4450 & 0.80 & 4550 & 1.07 & --2.32 & --2.35 \\ 
  S-6  & 4450 & 0.90 & 4390 & 0.92 & --2.45 & --2.50 \\
  S-8  & 4625 & 1.30 & 4475 & 0.95 & --2.55 & --2.61 \\
\enddata
\end{deluxetable}

\clearpage

\begin{deluxetable}{crrrrrrrr}
\tablenum{A7}
\tablecaption{Fe Abundances for Ten Giants in M71\label{tableA7}}
\tablewidth{0pt}
\tablehead{
\colhead{Star}              & 
\colhead{T$_{\rm eff}$}         &
\colhead{log~$g$}           &
\colhead{T$_{\rm eff}$}         &
\colhead{log~$g$}           &
\colhead{[Fe/H]$_{\rm II}$} &
\colhead{[Fe/H]$_{\rm II}$} &
\colhead{[Fe/H]$_{\rm I}$}  &
\colhead{[Fe/H]$_{\rm I}$}  \\
\colhead{}                  &
\colhead{orig}              &
\colhead{orig}              &
\colhead{rev}               &
\colhead{rev}               &
\colhead{orig}              &
\colhead{rev}               &
\colhead{orig}              &
\colhead{rev}               
}                      
\startdata
I-113 & 3950 & 0.70 &  3940 &  0.83 & --0.82 & --0.77 & --0.87 & --0.86 \\
I-46  & 4000 & 0.80 &  3980 &  0.80 & --0.86 & --0.86 & --0.79 & --0.79 \\
I-45  & 4050 & 0.80 &  4030 &  0.87 & --0.84 & --0.80 & --0.78 & --0.77 \\
A4    & 4100 & 0.80 &  4060 &  0.84 & --0.79 & --0.75 & --0.80 & --0.79 \\
I-77  & 4100 & 0.95 &  4040 &  0.83 & --0.78 & --0.75 & --0.80 & --0.84 \\
A9    & 4200 & 1.20 &  4200 &  1.24 & --0.88 & --0.86 & --0.87 & --0.87 \\
I     & 4200 & 1.00 &  4220 &  1.01 & --0.95 & --0.95 & --0.91 & --0.91 \\
I-53  & 4300 & 1.40 &  4175 &  1.22 & --0.89 & --0.88 & --0.81 & --0.82 \\
S     & 4300 & 1.25 &  4280 &  1.32 & --0.76 & --0.73 & --0.74 & --0.73 \\
I-21  & 4350 & 1.45 &  4480 &  1.46 & --0.81 & --0.71 & --0.75 & --0.81 \\
\enddata
\end{deluxetable}

\clearpage

\begin{deluxetable}{crrrrrrrr}
\tablenum{A8}
\tablecaption{Fe Abundances for Nine Giants in M10\label{tableA8}}
\tablewidth{0pt}
\tablehead{
\colhead{Star}              & 
\colhead{T$_{\rm eff}$}         &
\colhead{log~$g$}           &
\colhead{T$_{\rm eff}$}         &
\colhead{log~$g$}           &
\colhead{[Fe/H]$_{\rm II}$} &
\colhead{[Fe/H]$_{\rm II}$} &
\colhead{[Fe/H]$_{\rm I}$}  &
\colhead{[Fe/H]$_{\rm I}$}  \\
\colhead{}                  &
\colhead{orig}              &
\colhead{orig}              &
\colhead{rev}               &
\colhead{rev}               &
\colhead{orig}              &
\colhead{rev}               &
\colhead{orig}              &
\colhead{rev}               
}                      
\startdata
A-I-2    & 3975 & 0.00 & 4030 & 0.45 & --1.49 & --1.36 & --1.49 & --1.47 \\
A-III-21 & 4060 & 0.50 & 4120 & 0.54 & --1.54 & --1.58 & --1.51 & --1.64 \\
A-II-24  & 4050 & 0.10 & 4050 & 0.55 & --1.56 & --1.38 & --1.52 & --1.50 \\
A-III-16 & 4150 & 0.90 & 4225 & 0.77 & --1.54 & --1.65 & --1.54 & --1.49 \\
H-I-367  & 4135 & 0.60 & 4200 & 0.82 & --1.52 & --1.44 & --1.56 & --1.50 \\
H-I-15   & 4225 & 0.75 & 4300 & 0.92 & --1.52 & --1.56 & --1.54 & --1.48 \\
A-III-5  & 4400 & 1.20 & 4340 & 1.09 & --1.42 & --1.59 & --1.38 & --1.44 \\
A-I-60   & 4400 & 1.10 & 4475 & 1.29 & --1.53 & --1.49 & --1.50 & --1.48 \\
A-I-61   & 4650 & 1.20 & 4600 & 1.38 & --1.68 & --1.53 & --1.71 & --1.64 \\
\enddata
\end{deluxetable}

\clearpage

\begin{deluxetable}{crrrrrrl}
\tablenum{A9}
\tablecaption{ Fe Abundances for Four Giants in NGC 7006\label{tableA9}}
\tablewidth{0pt}
\tablehead{
\colhead{Star}              & 
\colhead{T$_{\rm eff}$}         &
\colhead{log~$g$}           &
\colhead{T$_{\rm eff}$}         &
\colhead{log~$g$}           &
\colhead{[Fe/H]$_{\rm II}$} &
\colhead{[Fe/H]$_{\rm I}$}  &
\colhead{Remark}            \\
\colhead{}                  &
\colhead{orig}              &
\colhead{orig}              &
\colhead{rev}               &
\colhead{rev}               &
\colhead{rev}               &
\colhead{rev}               &
\colhead{}
}                      
\startdata
  I-1    & 3900 & 0.10 & 3825 & 0.23 & --1.35 &  --1.64 &   0.5 wt \\
  II-103 & 4200 & 0.75 & 4175 & 0.76 & --1.56 &  --1.59 & \\
  II-46  & 4200 & 0.50 & 4210 & 0.86 & --1.42 &  --1.64 & \\
  II-18  & 4300 & 0.90 & 4275 & 0.88 & --1.53 &  --1.60 & \\
\enddata
\end{deluxetable}

\clearpage

\begin{deluxetable}{crrrcccc}
\tablenum{A10}
\tablecaption{Fe Abundances for 13 Giants in NGC 3201\label{tableA10}}
\tablewidth{0pt}
\tablehead{
\colhead{Star}              & 
\colhead{T$_{\rm eff}$}         &
\colhead{log~$g$}           &
\colhead{T$_{\rm eff}$}         &
\colhead{log~$g$}           &
\colhead{[Fe/H]\tablenotemark{(a)}} &
\colhead{[Fe/H]$_{\rm II}$} &
\colhead{[Fe/H]$_{\rm I}$}  \\
\colhead{Lee\#}             &
\colhead{orig}              &
\colhead{orig}              &
\colhead{rev}               &
\colhead{rev}               &
\colhead{orig}              &
\colhead{rev}               &
\colhead{rev}               
}                      
\startdata
1117 & 4000 & 0.2  & 4025 & 0.47 &  --1.47 &  --1.37 & --1.42 \\
1312 & 4000 & 0.0  & 3925 & 0.31 &  --1.45 &  --1.38 & --1.49 \\
1314 & 4150 & 0.4  & 4150 & 0.45 &  --1.58 &  --1.57 & --1.58 \\
1410 & 4250 & 0.7  & 4370 & 1.11 &  --1.64 &  --1.59 & --1.49 \\
2405 & 4250 & 1.3  & 4250 & 0.94 &  --1.43 &  --1.59 & --1.45 \\
3218 & 4150 & 0.7  & 4150 & 0.94 &  --1.60 &  --1.48 & --1.61 \\
3401 & 4250 & 0.9  & 4225 & 0.92 &  --1.47 &  --1.46 & --1.48 \\
3414 & 4500 & 0.8  & 4400 & 1.18 &  --1.46 &  --1.26 & --1.52 \\
3522 & 4350 & 0.8  & 4325 & 1.02 &  --1.57 &  --1.49 & --1.57 \\
3616 & 4250 & 1.2  & 4250 & 0.88 &  --1.44 &  --1.56 & --1.43 \\
4319 & 4250 & 1.0  & 4120 & 1.17 &  --1.54 &  --1.62 & --1.64 \\
4507 & 4250 & 0.7  & 4250 & 0.96 &  --1.52 &  --1.40 & --1.50 \\
4524 & 4150 &-0.1  & 4125 & 0.52 &  --1.73 &  --1.43 & --1.69 \\
\enddata
\tablenotetext{(a)}{Original values of Fe~{\sc i} and Fe~{\sc ii} 
differ no more than 0.01 \citep{gonzalez02}.}
\end{deluxetable}

\clearpage

\end{document}